\documentclass{article}

\usepackage[preprint]{neurips_2026}

\usepackage[utf8]{inputenc} 
\usepackage[T1]{fontenc}    
\usepackage{hyperref}       
\usepackage{url}            
\usepackage{booktabs}       
\usepackage{amsfonts}       
\usepackage{nicefrac}       
\usepackage{microtype}      
\usepackage{xcolor}         

\usepackage{subcaption}
\usepackage{graphicx}
\usepackage{amsmath} 
\usepackage{comment}

\title{\textit{Yeti}: A compact protein structure tokenizer for reconstruction and multi-modal generation}

\author{%
}

\author{
\begin{tabular}{ccc}
Nabin Giri & Steven Farrell & Kristofer E. Bouchard \\
\texttt{ngiri@lbl.gov} & \texttt{sfarrell@lbl.gov} & \texttt{kebouchard@lbl.gov} \\
\\ 
\multicolumn{3}{c}{Lawrence Berkeley National Laboratory} \\
\multicolumn{3}{c}{1 Cyclotron Road, Berkeley, CA 94720}
\end{tabular}
}

\begin{document}

\maketitle

\begin{abstract}
Multi-modal models that jointly reason over protein sequences, structures, and function annotations within a unified representation hold immense potential for integrating multi-modal data and generating new proteins with designed functional properties. To utilize transformer architectures, such models require a tokenizer that converts protein structure from continuous atomic coordinates into discrete representations suitable for scalable multi-modal training. 
The quality of such models are fundamentally upper-bounded by the fidelity and expressiveness of the underlying tokenized structure.  However, existing tokenizers prioritize reconstruction over generative abilities. To address these gaps, we introduce \textit{Yeti}, a simple and compact protein structure tokenizer based on look-up free quantization and trained end-to-end with a flow matching objective for multi-modal learning. Compared to existing models, \textit{Yeti} generally achieves the best codebook utilization and token diversity, and second best reconstruction accuracy (with 10x fewer parameters than ESM3) on diverse datasets. To validate \textit{Yeti}’s generative capability, we trained a compact multi-modal model jointly over its structure tokens and amino acid sequence entirely from scratch, with no pretrained initialization. The resulting multi-modal model generates plausible structures under unconditional co-generation of protein sequence and structures, achieving comparable results to 10x larger models. Together, these results demonstrate that \textit{Yeti} is a compact and expressive protein structure tokenizer suitable for training multi-modal models that co-generates highly plausible sequences and structures.
\end{abstract}

\vspace{-10pt}
\section{Introduction}
\vspace{-10pt}
Biological systems are among the most complex and multifaceted systems in nature, shaped by billions of years of evolution and governed by intertwined physical, chemical, and biological processes. Understanding and modeling such systems has long been a central challenge in computational biology, driving the development of innovative methods that balance accuracy, flexibility, and scalability. However, current methods struggle along three interconnected areas; first, biological data is inherently heterogeneous spanning biophysical constraints, biochemical annotations, and experimentally derived modalities such as function-annotated text, two-dimensional imaging, and three-dimensional volumetric data from cryo-electron microscopy and tomography with each modality carrying distinct statistical properties and noise characteristics \cite{cryo2structdata, cryodenoise_data}. Second, real biological datasets are incomplete by nature; modalities are frequently missing, measurements are noisy, and coverage across the structure and function space of proteins remains uneven. Third, biological processes are multiscale and coupled, ranging from atomic interactions to cellular behavior, yet most models operate at a single scale and modality without mechanisms to transfer information across levels. These challenges are compounded by the growing demand for models that are not only accurate but also steerable, interpretable, and sample-efficient, a requirement that become critical as we move toward unified multi-modal systems capable of modeling the full complexity of biological system.

Proteins sit at the center of this challenge. As the primary functional units of the cell, their function is determined not just by one-dimensional (1D) sequence alone but by three-dimensional (3D) atomic structure, and understanding that structure-function relationship requires representations that are both expressive and compatible with large-scale learning. Recent generative models have addressed this by treating proteins as 3D objects, utilizing complex architectures operating in SE(3) and diffusion processes in continuous coordinate space. A complementary and increasingly compelling alternative is to encode 3D atomic structure as discrete tokens, as in ESM3\cite{esm3}, DPLM2\cite{dplm2}, and SaProt\cite{saprot}, which frame protein modeling as masked or auto-regressive language modeling over structure vocabularies jointly with inherently discrete modalities such as sequence and functional annotations. An advantage of auto-regressive model is that it allows protein structure generation without predefined length which is desirable in a case where given an experimental cryo-EM density map we want to automatically generate a 3D atomic model without any information about number of residues present in the structure \cite{cryoem_review, cryo2struct}. A discrete tokenization of modalities allows the training of a multi-modal model which is scalable, modality-agnostic, and extensible to diverse data types within a unified framework.  In this spirit, we present a protein structure tokenizer, \textit{Yeti} (\textbf{Y}ielding \textbf{E}ncoded \textbf{T}okens for \textbf{I}ntermodality), that maps 3D atomic coordinates to discrete tokens suited for large-scale multi-modal learning, laying the groundwork for a universal molecular representation capable of capturing the full complexity of living systems.
\vspace{-5pt}
\section{Related Work}
\vspace{-10pt}
Protein structure tokenizers \cite{protein_tokenizer_benchmark_receipe, bio2token, foldtoken} work by converting the continuous 3D coordinates of proteins into a sequence of discrete tokens from a finite vocabulary utilizing vector quantization methods. We represent these structures using the input space $\mathbb{R}^{L \times 3}$. In this formulation, $L$ corresponds to the number of $C_\alpha$ residues present in the 3D atomic protein structure, and the three-dimensional vector denotes the coordinates of each $C_\alpha$ atom. 

The standard paradigm for training protein structure tokenizers typically follows a two-stage process. First, an encoder and codebook are optimized alongside a lightweight decoder. In the second stage, the encoder and codebook are frozen while a larger, more expressive decoder is trained to maximize reconstruction accuracy. Building on the foundations of AlphaFold2 \cite{jumper2021highly}, most architectures often rely on $SE(3)$-invariant components such as Invariant Point Attention (IPA) and $SE(3)$-invariant loss like Frame-Aligned Point Error (FAPE). ESM3 \cite{esm3} follows this two-stage approach with a VQ-VAE \cite{vqvae} quantizer and a 4,096-entry codebook. The first stage uses five supervised losses including backbone distance, backbone direction, and inverse folding, and the second stage trains a deeper decoder on all-atom geometric losses across PDB \cite{pdb}, AFDB, and ESMAtlas structures. The full tokenizer model contains $\approx$ 600M parameters. DPLM-2 \cite{dplm2} similarly follows the two-stage paradigm. Its tokenizer consists of a pretrained GVP-Transformer \cite{gvp_transformer} encoder which is frozen during the tokenizer training, a Lookup-Free Quantizer (LFQ) \cite{lfq}, and an IPA-based decoder from AlphaFold2. Reconstruction is supervised with FAPE, violation, and distogram losses inherited from AlphaFold2, along with a sequence prediction cross-entropy head. This tokenizer contains approximately 100M parameters. Kanzi \cite{kanzi} departs from the standard two-stage paradigm by using a flow-matching decoder over raw global coordinates, without $SE(3)$-invariant representations. It utilizes an asymmetric architecture (2-layer encoder, 8–12 layer decoder) with Finite Scalar Quantization (FSQ) \cite{fsq} and a relatively small codebook ($\approx$ 1,000 entries). The publicly available kanzi model is limited to proteins with $\leq 256$ residues, and its generative capability is demonstrated via an auto-regressive model trained on its tokenized structures. However, as a single-modality approach, its multi-modal generative potential remains underexplored. Additionally, its use of sliding window attention can induce token repetition during encoding, which degrades downstream generative task \cite{dilip2026adaptive, scaling_codebook}. A common assumption is that better reconstruction translates to better generation. These objectives are not inherently aligned~\cite{lfq, scaling_visual_tokens_learnings, yao2025reconstruction} and a tokenizer can achieve high reconstruction accuracy while producing tokens poorly suited for generative modeling task, and vice versa. Despite this, most evaluation protocols treat reconstruction quality as the primary proxy for tokenizer quality. We suggest this is insufficient, and advocate evaluating tokenizers on both reconstruction accuracy and support for high-quality generation under multi-modal model paradigm.

To address these gaps, we present \textit{Yeti}, a single-stage protein structure tokenizer built on the StripedHyena~\cite{stripehyena} architecture and trained end-to-end under a flow-matching objective. \textit{Yeti} shows high codebook utilization across an 8,192-entry codebook over diverse proteins (L $\leq$ 512) without any pretrained sequence backbone. Protein structure reconstruction using \textit{Yeti} tokens is competitive with models having $10\times$ as many parameters. We trained a Masked Diffusion Model jointly over \textit{Yeti}'s structure tokens and sequence entirely from scratch under limited data and compute, and we demonstrate a strong \textit{unconditional sequence–structure co-generation}, validating that \textit{Yeti}'s tokens encode rich information to support training of multi-modal generative models. We further analyze the decoding trajectory and identify a hierarchical emergence of secondary structure across flow steps, culminating in a \textit{late commitment} phenomenon where global protein topology consolidates only in the final decoding steps.

\begin{figure}[t]
  \centering
   \includegraphics[width=0.9\textwidth]{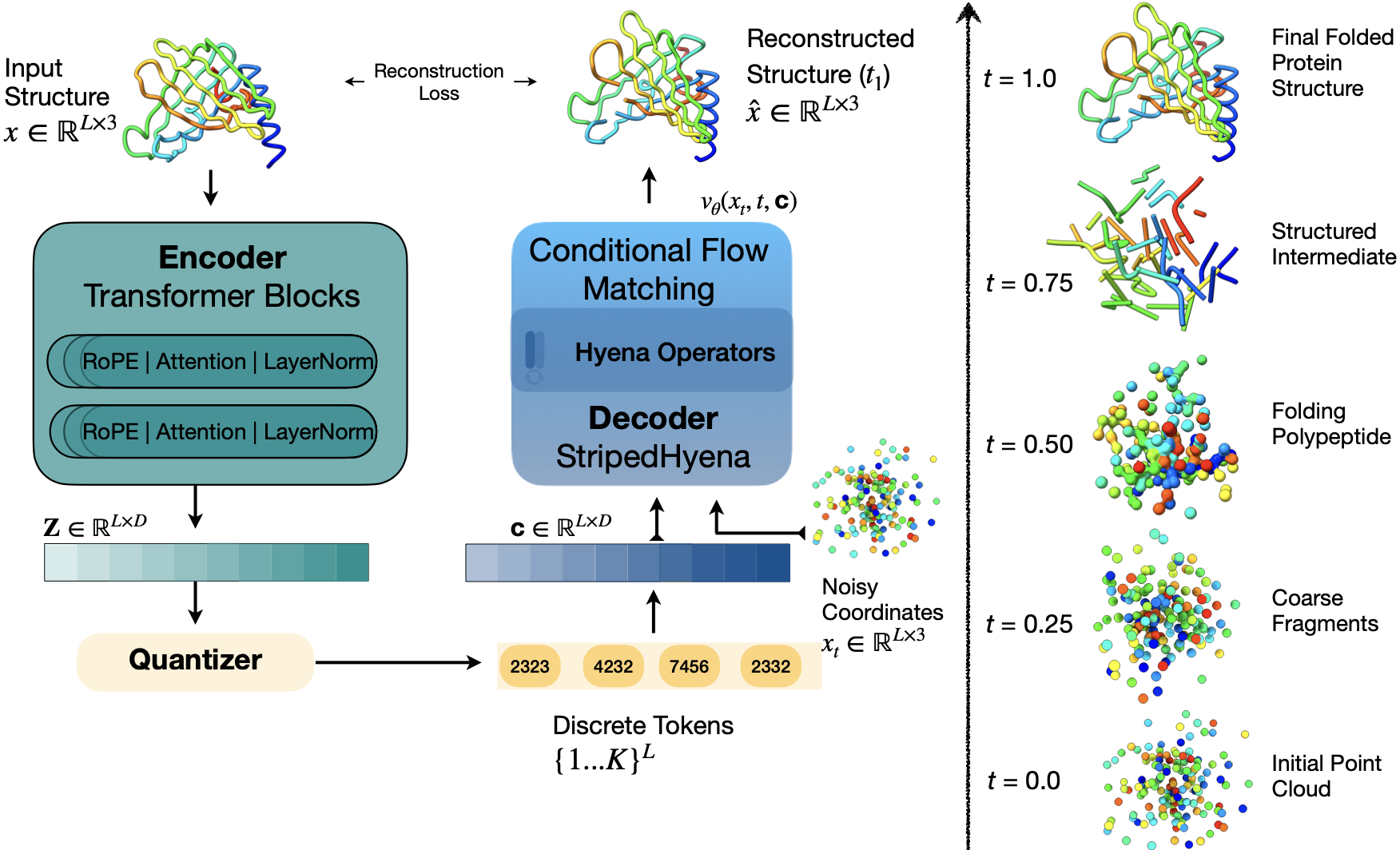}
     \caption{\textbf{\textit{Yeti} Architecture Overview.} \textbf{(Left)} A Transformer encoder maps input protein structures $x \in \mathbb{R}^{L \times 3}$ into continuous latent embeddings $\mathbf{Z} \in \mathbb{R}^{L \times D}$. LFQ produces quantized representations which are subsequently mapped into discrete tokens. \textbf{(Center)} The discrete tokens are projected into conditioning embeddings $\mathbf{c} \in \mathbb{R}^{L \times D}$ and provided to a StripedHyena decoder with Hyena operators. The decoder learns a conditional velocity field $\mathbf{v}_{\theta}(\mathbf{x}_t, t, \mathbf{c})$ to reconstruct protein structures from noisy coordinates. \textbf{(Right)} The learned generative trajectory progressively transforms an initial noisy point cloud $t=0.0$ into a folded protein structure $t=1.0$.}
     \label{fig:structure_tokenizer_arc}\hfill
\end{figure}
\vspace{-5pt}
\section{Methods}
\vspace{-10pt}
\paragraph{Tokenization.}
The high-level overview of \textit{Yeti}'s architecture is shown in \textbf{Figure \ref{fig:structure_tokenizer_arc}}. Given a clean protein structure with mean-centered coordinates \(x\), the encoder processes the input to generate latent embeddings $\mathbf{Z} = E(x) \in \mathbb{R}^{L \times D}$, where \(L\) denotes the protein length. The encoder consists of a stack of Transformer~\cite{vaswani2017attention} blocks utilizing multi-head attention with rotary positional encoding \cite{roformer}. The latent embeddings are further projected into a quantization space with dimensionality \(D = \log_2 K\), where \(K\) denotes the codebook size. Each embedding vector $\mathbf{z} \in \mathbb{R}^{D}$ is then passed through the Lookup-Free Quantizer (LFQ) \cite{lfq} \(q\). The LFQ latent space is defined as the Cartesian product $\mathcal{C} = \prod_{i=1}^{\log_2 K} C_i$, where each sub-codebook is binary, $C_i = \{-1, +1\}$. Given \(\mathbf{z} \in \mathbb{R}^{\log_2 K}\),a feature vector, each dimension of the quantized representation \(q(\mathbf{z})\) is obtained as $ q(z_i) = C_{i,j}$, where $ 
j = \arg\min_k |z_i - C_{i,k}|$, where \(C_{i,j}\) denotes the \(j\)-th element of \(C_i\). Since \(C_i = \{-1, +1\}\), the quantization reduces to a sign operation: $q(z_i)
=
\mathrm{sign}(z_i)
=
-\mathbf{1}\{z_i \le 0\}
+
\mathbf{1}\{z_i > 0\}.
$ The binary quantized representation is mapped to a discrete token index by:
\begin{equation}
\mathrm{Index}(\mathbf{z})
=
\sum_{i=1}^{\log_2 K}
2^{i-1}
\mathbf{1}\{z_i > 0\},
\label{eq:lfq_index}
\end{equation}
yielding a vocabulary size of \(K = 2^D\).

\vspace{-5pt}
\paragraph{Flow Matching.}
Flow matching \cite{lipman2022flow} interpolates between a source distribution $p_{0}$ (a Gaussian) and a target distribution $p_{data}$ by integrating the ordinary differential equation (ODE) $d$x$_{t}$ = v$_{\theta}$(x$_{t},t)dt$ where the learned vector field is defined as v$_{\theta}$(x$_{t},t) : \mathbb{R}^{d} \times $[0,1] $\to \mathbb{R}^{d}$. Since the true vector field generating the data distribution is generally unknown, we use conditional flow matching, which constructs a conditional probability path between noise samples x$_{0} \sim p_{0}$ and data samples x$_{1} \sim p_{data}$. A general probability path is defined as x$_{t} = \alpha_{t}x_{1} + \sigma_{t}\epsilon$, with $\epsilon \sim \mathcal{N}(0,1)$ which induces the conditional vector field \texttt{u}(x$_{t}|x_{0}, x_{1})$ = $\dot{\alpha}_{t}x_{1} + \dot{\sigma}_{t}x_{0}$, where the dot denotes the time derivative. For the standard linear interpolation path, x$_{t} = (1 - t)x_{0} + tx_{1}$ the target vectory field simplifies to \texttt{u}(x$_{t}|x_{0}, x_{1}) = x_{1} - x_{0}$. The decoder predicts the conditional velocity field  $\mathbf{v}_\theta(\mathbf{x}_t, t, \mathbf{c})$, where $\mathbf{x}_t$ denotes the intermediate structure at time $t$ and $\mathbf{c}$ represents the embeddings produced by the post quantizer projection layer following the LFQ module.
\vspace{-5pt}
\paragraph{Loss Function.}
Let $x_1$ denote a clean data sample (protein structure) and $x_0$ denote a sample from the base distribution (Gaussian noise with zero mean). We define the time-dependent probability path as $x_t = (1-t)x_0 + t x_1 $, where $t \sim \mathcal{U}[0,1]$ is the time step. The training objective is based on the loss function:

\begin{equation}
    \mathcal{L}_{FM}(\theta) = \mathbb{E}_{t, x_0, x_1} \left[ \| \mathbf{v}_\theta(x_t, t, \mathbf{c}) - (x_1 - x_0) \|^2 \right],
\end{equation}
We add an entropy penalty to LFQ during training to encourage codebook utilization: 
\begin{equation}
\mathcal{L}_{\mathrm{entropy}}
=
\mathbb{E}
\left[
H\!\left(q(\mathbf{z})\right)
\right]
-
H\!\left(
\mathbb{E}
\left[
q(\mathbf{z})
\right]
\right).
\label{eq:entropy_loss}
\end{equation}

The overall training objective is:
\begin{equation}
\mathcal{L}
=
\mathcal{L}_{\mathrm{FM}}
+
\lambda
\mathcal{L}_{\mathrm{entropy}},
\label{eq:total_loss}
\end{equation}

where \(\lambda\) controls the strength of entropy regularization, we set $\lambda$ to 0.1. Here, \(\mathbf{c}\) denotes the quantized conditioning embedding generated from the LFQ tokens, and $v_\theta(x_t, t, \mathbf{c})$ is the conditional velocity field predicted by the StripedHyena decoder. The model learns to approximate the optimal transport direction \((\mathbf{x}_1 - \mathbf{x}_0)\) between the Gaussian prior and the protein structure distribution.

We perform a series of ablation study to identify the optimal configuration for \textit{Yeti} and provide details in Appendix Section \ref{sec:model_scaling}. The training implementation details are provided in Appendix Section \ref{sec:structure_tokenizer}.

\vspace{-5pt}
\paragraph{Inference.}
To improve the protein structure quality, during the inference phase we utilize a classifier-free guidance \cite{cfg} technique. Since we trained the decoder to jointly learn the conditional velocity field $\mathbf{v}_\theta(\mathbf{x}_t, t, \mathbf{c})$ and the unconditional prior $\mathbf{v}_\theta(\mathbf{x}_t, t, \emptyset)$ by randomly zeroing out the token embeddings with a probability of $0.15$ during training, the guided velocity $\hat{\mathbf{v}}_t$ is computed as:

\begin{equation}
\hat{\mathbf{v}}_\theta(\mathbf{x}_t, t, \mathbf{c}) = \mathbf{v}_\theta(\mathbf{x}_t, t, \emptyset) + w \big( \mathbf{v}_\theta(\mathbf{x}_t, t, \mathbf{c}) - \mathbf{v}_\theta(\mathbf{x}_t, t, \emptyset) \big)
\label{eq:cfg}
\end{equation}

where $w$$\geq$1 is the guidance scale. We ran a hyperparameter sweep over $w \in \{1.4, 1.8, \dots, 4.0\}$ using 5,000 test proteins from the CATH test set (details of the dataset is provided in Appendix \ref{sec:test_dataset_location}) to determine the optimal guidance strength. We find $w=3.8$ yields best results on global topology and local geometric accuracy as measured by TM-score and RMSD, respectively. Notably, while the unguided model ($w=1.0$) demonstrates a strong structure accuracy with a mean RMSD of 1.50 \AA \ and TMscore of 0.93, the guided decoding at $w=3.8$ acts as a refinement, reducing the RMSD to 1.25\AA \ and increasing TMscore to 0.95.\\

\vspace{-5pt}
\paragraph{Datasets.} The first dataset we prepared for training \textit{Yeti} contains 614,718 single-chain protein structures extracted from the AlphaFold Protein Structure Database \cite{jumper2021highly}, clustered at 50\% sequence similarity using Foldseek \cite{foldseek} to reduce redundancy, which we refer to as $\mathcal{D}_{\textit{AFDB}}$. Additionally, in the later stage of \textit{Yeti} training, we further continue the training of the model with a subset of ESMAtlas dataset for few steps (yielded better results), which contains 2,085,441 single-chain high quality protein structures clustered at 30\% sequence similarity from MGnify, which we refer to as $\mathcal{D}_{\textit{ESM}}$. The reason for the second dataset is presented in Appendix Section \ref{sec:continued_training_dataset}. We refer combination of both these dataset as $\mathcal{D}_{\textit{A+E}}$. Appendix \textbf{Figure \ref{fig:train_data_composition}} shows the training/validation data statistics, more detail on dataset is available in Appendix Section \ref{sec:dataset_stats}. Our testing dataset contains CATH, CASP14, CASP15, and CAMEO dataset.

\vspace{-5pt}
\section{Results} \label{sec:results}
\vspace{-10pt}

To evaluate \textit{Yeti}, we compare it against representative tokenizers from the three dominant architecture families, namely, Vector Quantised Variational AutoEncoder (VQ-VAE) \cite{vqvae}, Lookup-Free Quantization (LFQ) \cite{lfq}, and Finite Scalar Quantization (FSQ) \cite{fsq}. Specifically, we select ESM3 (VQ-VAE) \cite{esm3}, DPLM-2 (LFQ) \cite{dplm2}, and Kanzi (FSQ) \cite{kanzi} as our primary baselines. For Kanzi, we evaluate the publicly available 44.1M-parameter checkpoint, which exceeds the 11M and 30M configurations reported in the original paper; we use it as-is to reflect the best publicly accessible variant. For fair comparison with Kanzi \cite{kanzi}, we restrict our evaluation to sequences within the range $50 \le L \le 256$, as this corresponds to the maximum supported length of the publicly available Kanzi checkpoint. We note that while this evaluation is bounded by the baseline's constraints, \textit{Yeti} is capable of processing up to $L = 512$.
\vspace{-5pt}
\subsection{Codebook Analysis}
\vspace{-10pt}
\textbf{Q1}. \textit{How effectively does Yeti utilize its discrete codebook?}\\
We evaluate the codebook dynamics of \textit{Yeti} by analyzing the entropy ($H$) and perplexity ($Px$) of the empirical token distribution across multiple test datasets. For a codebook of size $K=8192$, the theoretical maximum entropy under a uniform distribution is $H_{\max} = \ln(K) \approx 9.01$. As detailed in \textbf{Table~\ref{tab:codebook_detailed}} (metrics are described in Appendix \ref{sec:eval_metrics}), \textit{Yeti} consistently achieves high entropy and perplexity, significantly outperforming the VQ-VAE baseline (ESM3) and the FSQ baseline (Kanzi), while remaining highly competitive with the LFQ-based DPLM-2 (pretrained encoder). On CASP15, \textit{Yeti} achieves a perplexity of $3097$, a $\approx 30\%$ improvement over DPLM-2 ($Px=2375$) and more than double that of ESM3 ($Px=1446$). This is a good result as larger codebooks are harder to fully utilize because each additional entry must earn representation through training. Since \textit{Yeti} achieves high utilization of an 8,192-entry codebook, it shows that LFQ combined with end-to-end training effectively helps to produce a vocabulary where every structure token is semantically distinct and actively used. Despite having the same LFQ framework as DPLM-2, our approach achieves better codebook utilization. We attribute this gain to the differences in training paradigms, DPLM-2 utilizes a two-stage tokenizer training (encoder frozen) process whereas we train end-to-end. On the larger CATH dataset, \textit{Yeti} utilizes $\approx 66\%$ of its total capacity ($Px=5414$), but DPLM2 out performs here.
\begin{table}[t]
  \caption{\textbf{Codebook Utilization and Token Diversity.} Detailed comparison of entropy ($H$), perplexity ($Px$), and intra-structure diversity ($\bar{U}$) across test datasets. \textbf{Bold} and \underline{underline} denote best and second-best. \textbf{CB}: Codebook size; \textbf{$H$}: Entropy; \textbf{$Px$}: Perplexity ($\exp(H)$); \textbf{$\bar{U}$}: Average intra-structure diversity (\%). Note that $H$ and $Px$ are bounded by $\log(\text{CB})$ and CB respectively. All metrics measured on structures with $50 < L \leq 256$. Bold represents best score.}
  \label{tab:codebook_detailed}
  \centering
  \scriptsize
  \setlength{\tabcolsep}{6pt} 
  \begin{tabular}{l r | ccc | ccc | ccc | ccc}
    \toprule
    & & \multicolumn{3}{c}{\textbf{CAMEO} ($n=104$)} & \multicolumn{3}{c}{\textbf{CASP14} ($n=34$)} & \multicolumn{3}{c}{\textbf{CASP15} ($n=34$)} & \multicolumn{3}{c}{\textbf{CATH} ($n=11,752$)} \\
    \cmidrule(lr){3-5} \cmidrule(lr){6-8} \cmidrule(lr){9-11} \cmidrule(lr){12-14}
    \textbf{Model} & \textbf{CB} & $H\uparrow$ & $Px\uparrow$ & $\bar{U}\uparrow$ & $H\uparrow$ & $Px\uparrow$ & $\bar{U}\uparrow$ & $H\uparrow$ & $Px\uparrow$ & $\bar{U}\uparrow$ & $H\uparrow$ & $Px\uparrow$ & $\bar{U}\uparrow$ \\
    \midrule
    ESM3 \cite{esm3}    & 4,096 & 7.83 & 2519 & 93.6 & 7.56 & 1924 & 95.3 & 7.27 & 1446 & 88.2 & 7.96 & 2887 & 95.0 \\
    DPLM-2 \cite{dplm2}  & 8,192 & 8.37 & 4340 & 96.4 & 7.90 & 2708 & 97.5 & 7.77 & 2375 & 93.5 & \textbf{8.63} & \textbf{5584} & 97.1 \\
    Kanzi \cite{kanzi}   & 1,000 & 6.40 & 605 & 87.0 & 6.37 & 586 & 89.0 & 6.26 & 526 & 82.4 & 6.44 & 627 & 89.7 \\
    \midrule
    \textbf{Yeti} (Ours) & 8,192 & \textbf{8.40} & \textbf{4461} & \textbf{99.0} & \textbf{7.93} & \textbf{2779} & \textbf{99.0} & \textbf{8.04} & \textbf{3097} & \textbf{98.0} & {8.59} & 5414 & \textbf{98.8} \\
    \bottomrule
  \end{tabular}
\end{table}

A standout characteristic of \textit{Yeti} is the average intra-structure diversity ($\bar{U}$). Unlike global perplexity, $\bar{U}$ measures the variety of unique tokens used within a single protein structure, serving as a proxy for the information density of the latent representation. \textit{Yeti} nearly saturates this metric, achieving up to $99.0\%$ intra-structure diversity. This substantially exceeds ESM3 ($93.6\%$) and Kanzi ($87.0\%$), implying that \textit{Yeti} assigns highly distinct and non-redundant tokens to residues of individual proteins. Finally, we note that full codebook utilization (100\%) is neither expected nor theoretically desirable for protein structures. A well-structured discrete representation should naturally concentrate mass on common structure patterns (e.g., standard $\alpha$-helices and $\beta$-sheets) while reserving capacity for rarer loop regions or motifs. Consequently, some tokens may remain inactive on a given test set, especially when the dataset does not span the full diversity of protein structures, and thus incomplete utilization does not indicate inefficiency or collapse. The fact that \textit{Yeti} achieves high, but not total, utilization suggests it has successfully learned to compress the protein residues into discrete tokens. 

\vspace{-5pt}
\subsection{Reconstruction Capabilities}
\begin{figure}[t]
  \centering
   \includegraphics[width=\textwidth]{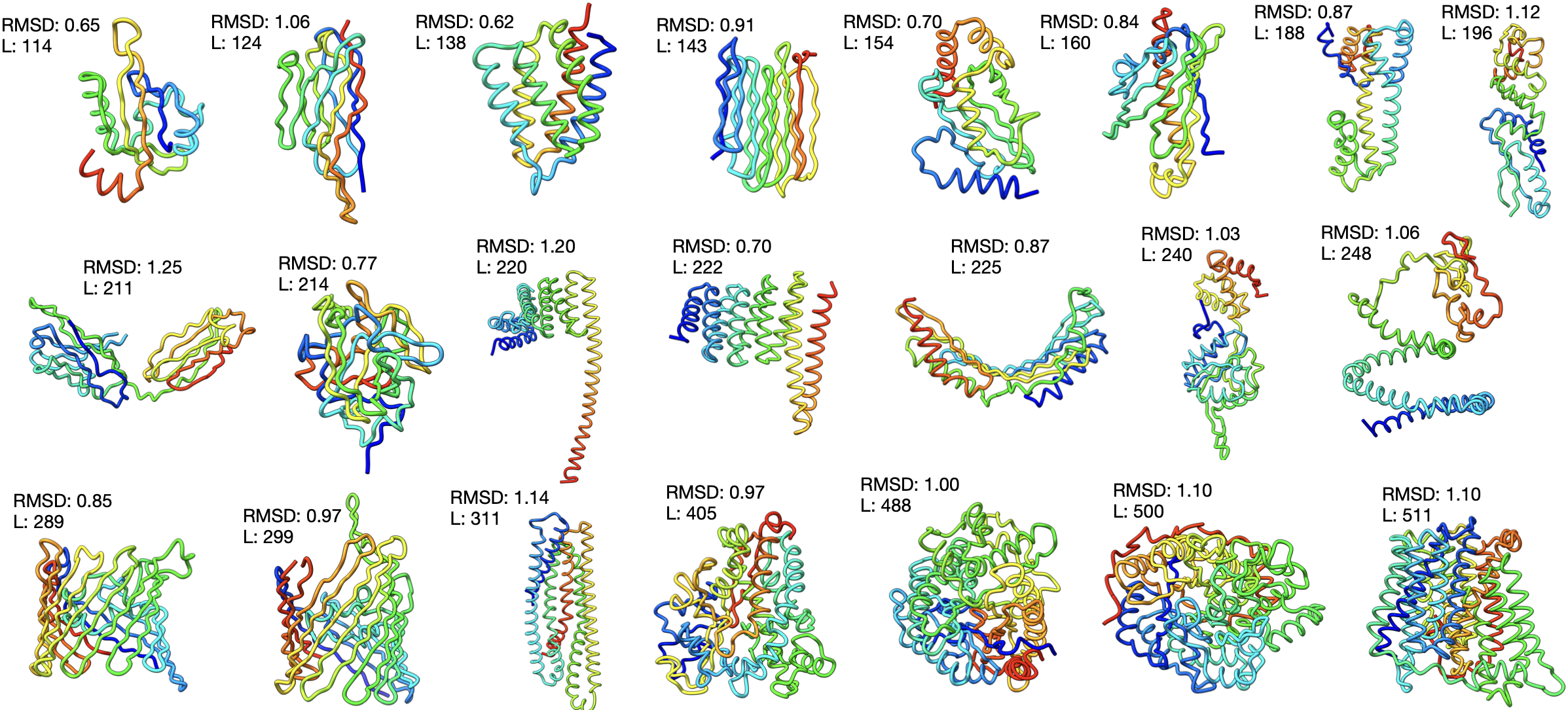}
     \caption{\textbf{Representative structure reconstructions of test proteins} Length ranging from $114 \leq L \leq 511$ residues; root-mean-square deviation (RMSD) values for each structure are included.}
     \label{fig:reconstructed_protein}\hfill
\end{figure}

\vspace{-10pt}
\textbf{Q2}. \textit{How accurately does \textit{Yeti} reconstruct protein structures from its discrete representations?}  \\
To evaluate how well \textit{Yeti}-based tokens enable protein structure reconstruction, we compared \textit{Yeti} against three state-of-the-art baselines relevant to our work, ESM3 \cite{esm3}, DPLM-2 \cite{dplm2}, and Kanzi \cite{kanzi}. We present the reconstruction results in \textbf{Table \ref{tab:reconstruction}}. We found that \textit{Yeti} achieved TM-score 0.96 on CAMEO, 0.97 on CASP14, 0.95 in CASP15 and CATH. This indicates that \textit{Yeti} consistently recovers correct protein global topology. \textit{Yeti} shows best reconstruction RMSD on the CASP15 test set and competitive performance on CAMEO and CASP14. Across all benchmarks, we observe that our TM-score advantage often exceeds our RMSD performance, we attribute this to the nature of the flow-based decoder. Unlike deterministic decoders that rely on IPA module \cite{jumper2021highly} that regress to a single coordinate set, our flow-based approach samples from a conditional distribution. Consequently, individual samples may show some variance in local atomic positions which penalizes RMSD while consistently recovering the correct global topology. This property can be desirable feature for downstream multi-conformation generation (different samples explore the local conformational space around the correct fold) but a limitation for the single-sample RMSD metric used in reconstruction evaluation. Representative examples of reconstructed proteins with varying RMSD values from test dataset from \textbf{Table \ref{tab:reconstruction}} are shown in \textbf{Figure \ref{fig:reconstructed_protein}}.

\textit{Yeti} shows significant parameter efficiency, achieving performance comparable to ESM3 (648M parameters, $\approx$ 10$\times$ larger) and DPLM-2 (118M parameters, $\approx$ 2$\times$ larger) with only 62.5M parameters. The primary advantage of ESM3’s scale is evident in its RMSD metrics, with low RMSD across the board. This represents a tradeoff between scale and efficiency as \textit{Yeti}'s design supports strong tokenization without requiring large-scale pretraining, while allowing an easy scalable architecture.

\begin{table}[t]
  \caption{\textbf{Protein Structure Reconstruction.} Comparison across diverse datasets for proteins ($50 < L \leq 256$). \textbf{Bold} and \underline{underline} denote best and second-best, respectively. RMSD ($\downarrow$, \AA): root mean square deviation of reconstructed vs.\ known C$\alpha$ coordinates.
  TM ($\uparrow$): TM-score measuring global fold similarity.}
  \label{tab:reconstruction}
  \centering
  \scriptsize
  \setlength{\tabcolsep}{3.5pt} 
  \begin{tabular}{l c | cc | cc | cc | cc}
    \toprule
    & & \multicolumn{2}{c}{\textbf{CAMEO} ($n=104$)} & \multicolumn{2}{c}{\textbf{CASP14} ($n=34$)} & \multicolumn{2}{c}{\textbf{CASP15} ($n=34$)} & \multicolumn{2}{c}{\textbf{CATH} ($n=11,752$)} \\
    \cmidrule(lr){3-4} \cmidrule(lr){5-6} \cmidrule(lr){7-8} \cmidrule(lr){9-10}
    \textbf{Model} & \textbf{Params} & RMSD$\downarrow$ & TM$\uparrow$ & RMSD$\downarrow$ & TM$\uparrow$ & RMSD$\downarrow$ & TM$\uparrow$ & RMSD$\downarrow$ & TM$\uparrow$ \\
    \midrule
    ESM3 \cite{esm3} & 648M & \textbf{0.87} $\pm$ 1.54 & \textbf{0.97} $\pm$ 0.06 & \textbf{0.58}  $\pm$ 0.43 & \textbf{0.98} $\pm$ 0.02 & \underline{1.38}  $\pm$ 1.49 & 0.93 $\pm$ 0.11 & \underline{1.07}  $\pm$ 1.71 & \textbf{0.96}  $\pm$ 0.07 \\
    DPLM-2 \cite{dplm2} & 118M & 1.65  $\pm$ 1.61 & 0.92  $\pm$ 0.07 & 1.57  $\pm$ 1.34 & 0.92  $\pm$ 0.07 & 4.87  $\pm$ 7.17 & 0.81  $\pm$ 0.20 & 1.55 $\pm$ 1.19 & 0.91 $\pm$ 0.07 \\
    Kanzi \cite{kanzi} & 44.1M & 1.06 $\pm$ 0.34 & 0.95 $\pm$ 0.02 & 0.97 $\pm$ 0.23 & 0.95  $\pm$ 0.01 & 1.49 $\pm$ 1.11 & 0.93  $\pm$ 0.06 & \textbf{1.07}  $\pm$ 0.46 & 0.94 $\pm$ 0.03 \\
    \midrule
    \textbf{\textit{Yeti}} (Ours) & 62.5M & \underline{1.05}  $\pm$ 0.82 & \underline{0.96}  $\pm$ 0.03 & \underline{0.81} $\pm$ 0.20 & \underline{0.97}  $\pm$ 0.01 & \textbf{1.27}  $\pm$ 0.81 & \textbf{0.95} $\pm$ 0.05 & 1.29  $\pm$ 1.20 & \underline{0.95}  $\pm$ 0.04 \\
    \bottomrule
  \end{tabular}
\end{table}

\vspace{-5pt}
\subsection{Generation Capabilities}
\vspace{-10pt}
\textbf{Q3.} \textit{Can Yeti tokens, under multi-modal training, support coherent and diverse unconditional co-generation of protein sequence and structure?} \\
We address the challenging task of jointly generating atomic structures and amino acid sequences. If successful, this could unlock fine-grained control over functional sites and allow protein design tasks, such as atomistic motif scaffolding. Towards this, we trained a Masked Diffusion Model (MDM) over an absorbing state \cite{austin2021structured, yu2506discrete} with Transformer architecture, jointly over \textit{Yeti}'s structure tokens ($x$) and amino acid sequences ($s$). We utilize the $\mathcal{D}_{\textit{A+E}}$ dataset. The unconditional co-generation task ($p_{\theta}(s, x)$) directly explores whether the model has learned the intrinsic coupling between amino acid sequence and three-dimensional atomic structure. Conventionally, this is addressed through cascaded pipelines, where either structure is generated first and a separate model is used to predict the sequence (e.g., ProteinMPNN \cite{mpnn}), or sequence is generated first and folded into structure (e.g., ESMFold \cite{esmfold}, AlphaFold \cite{jumper2021highly}), for more details see Appendix \ref{sec:multimodal_protein_models}. In contrast, we generate both modalities simultaneously in a single unified pass.

We found that a compact 224M-parameter model trained entirely from scratch on just 2.6M proteins based on \textit{Yeti} tokens, successfully learns a coherent joint distribution over protein sequence and structure. We achieved this at a fraction of the cost of existing approaches which is roughly 1/1,000th the training data of ESM3-Open~\cite{esm3} (2.6M vs.\ 2.78B sequences for a 1.4B-parameter model), and one-third the compute of DPLM-2~\cite{dplm2} (384 vs.\ 1{,}152 A100 GPU-hours). To the best of our knowledge, the flagship models like ESM3 and DPLM-2 are the only token-based methods publicly available with co-generation abilities \cite{co-gen-review}. We also note that it comes with caveats; DPLM-2 warm-starts from a 650M-parameter sequence language model, DPLM ~\cite{dplm}, pretrained on 45M UniRef50 sequences, then fine-tunes on 200K curated PDB \cite{pdb} and SwissProt structures, ESM3 runs its co-generation in a cascaded pipeline by first generating sequence then structure. In contrast, our model does not require pretrained model like DPLM2 and unlike ESM3 our model generates sequence-structure simultaneously.

\begin{figure}[t]
     \centering
        \begin{subfigure}[b]{0.19\textwidth}
         \centering
         \includegraphics[height=2.5cm]{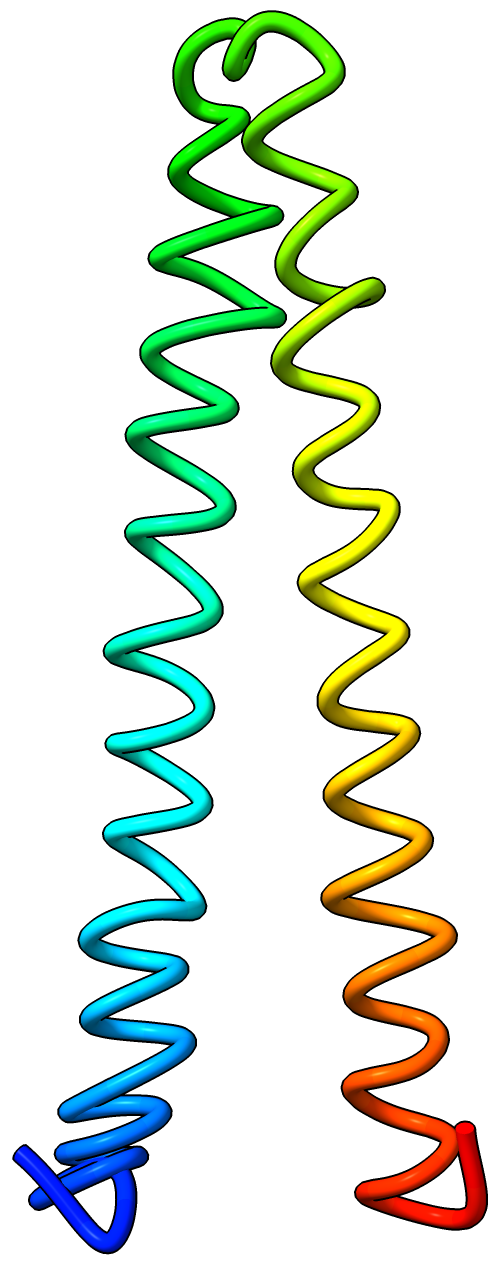}
         \caption{Length: 100.}
         \label{fig:uncond_100}
     \end{subfigure}
     \hfill
     \begin{subfigure}[b]{0.19\textwidth}
         \centering
         \includegraphics[height=2.5cm]{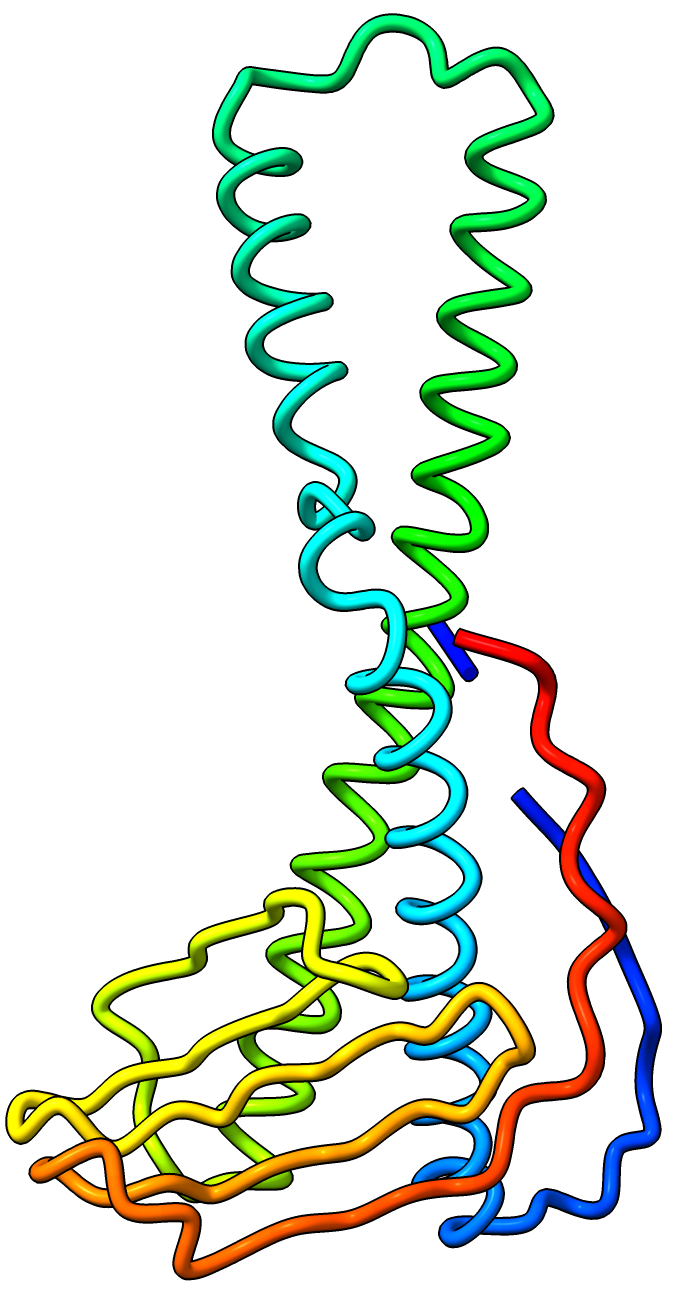}
         \caption{Length: 200.}
         \label{fig:uncond_200}
     \end{subfigure}
     \hfill
     \begin{subfigure}[b]{0.19\textwidth}
         \centering
         \includegraphics[height=2.5cm]{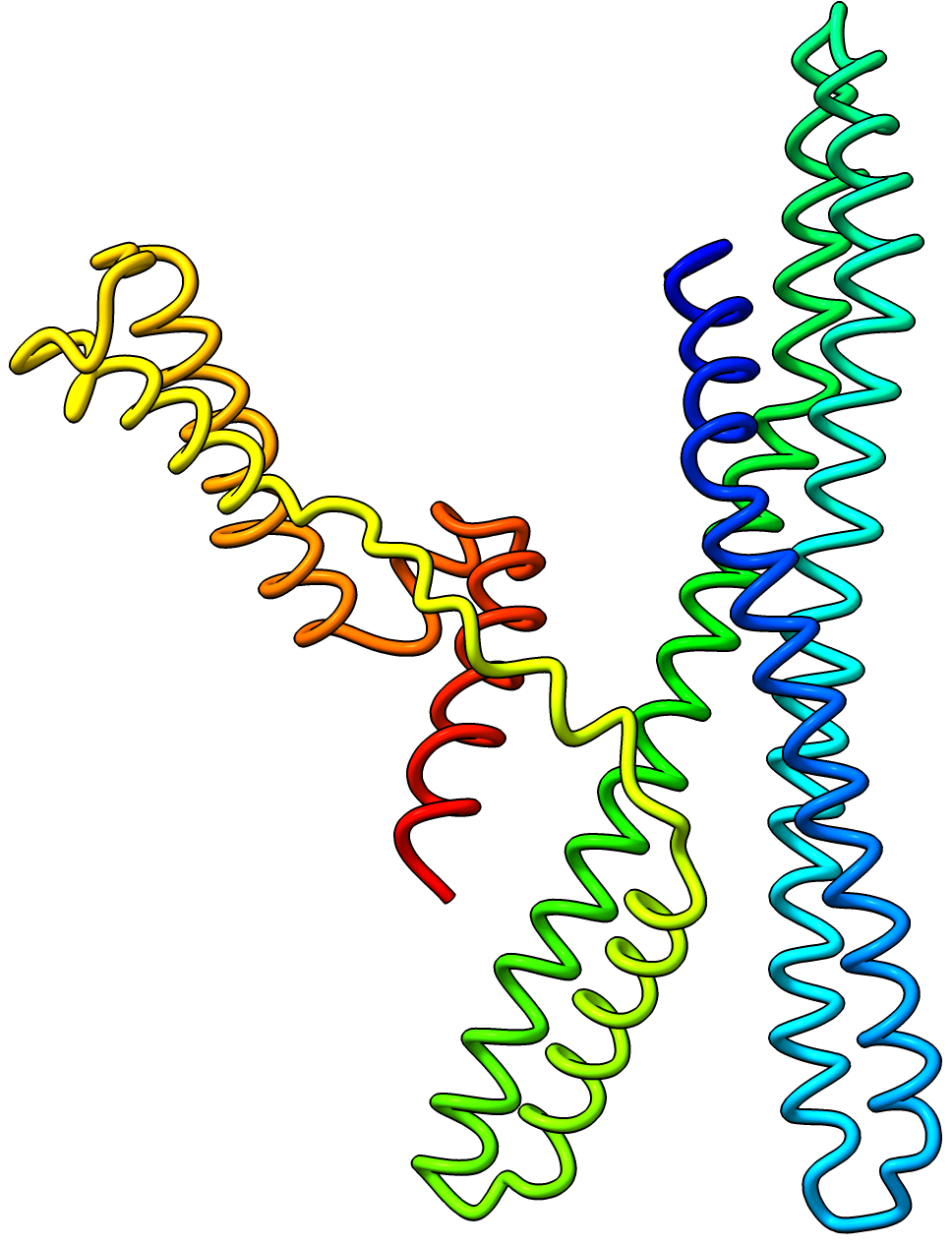}
         \caption{Length: 300.}
         \label{fig:uncond_300}
     \end{subfigure}
     \hfill
     \begin{subfigure}[b]{0.19\textwidth}
         \centering
         \includegraphics[height=2.5cm]{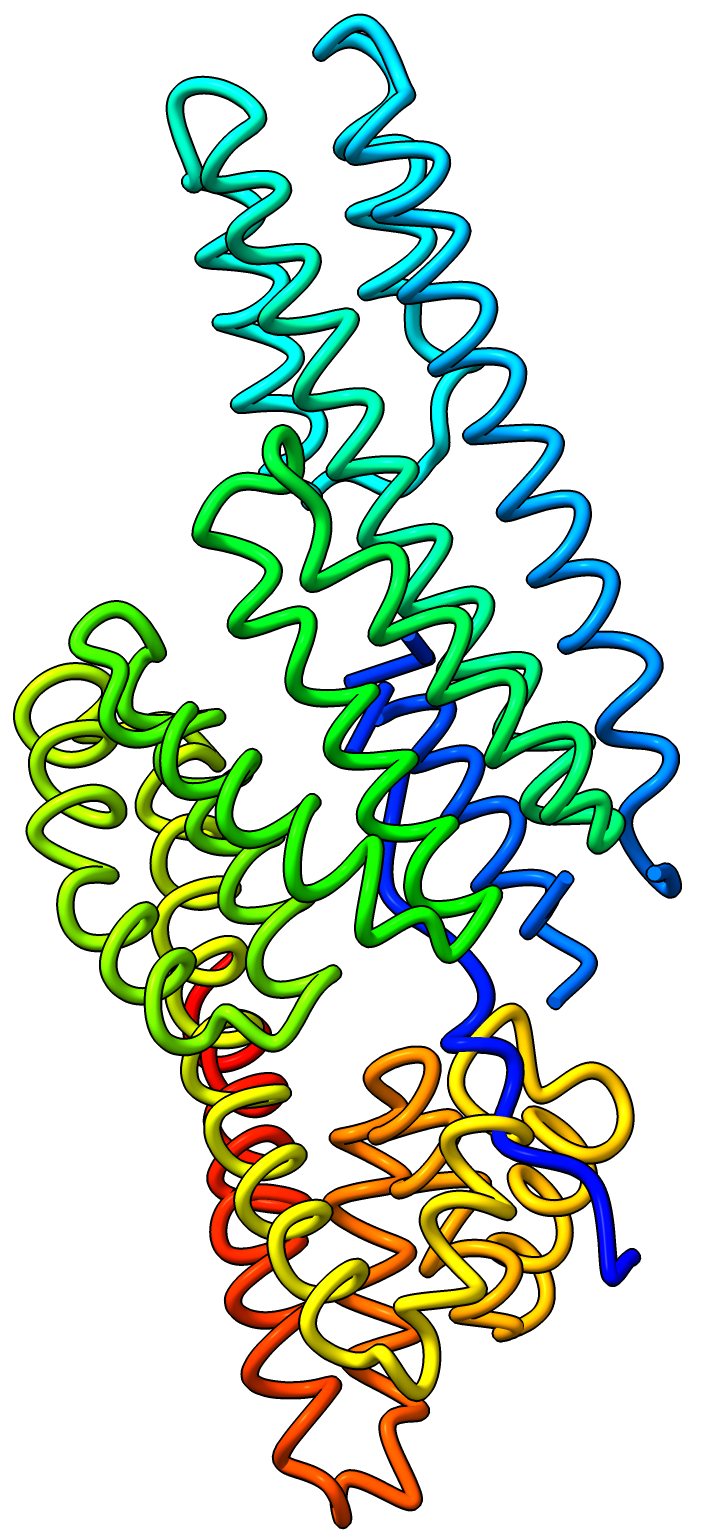}
         \caption{Length: 400.}
         \label{fig:uncond_400}
     \end{subfigure}
     \hfill
     \begin{subfigure}[b]{0.19\textwidth}
         \centering
         \includegraphics[height=2.5cm]{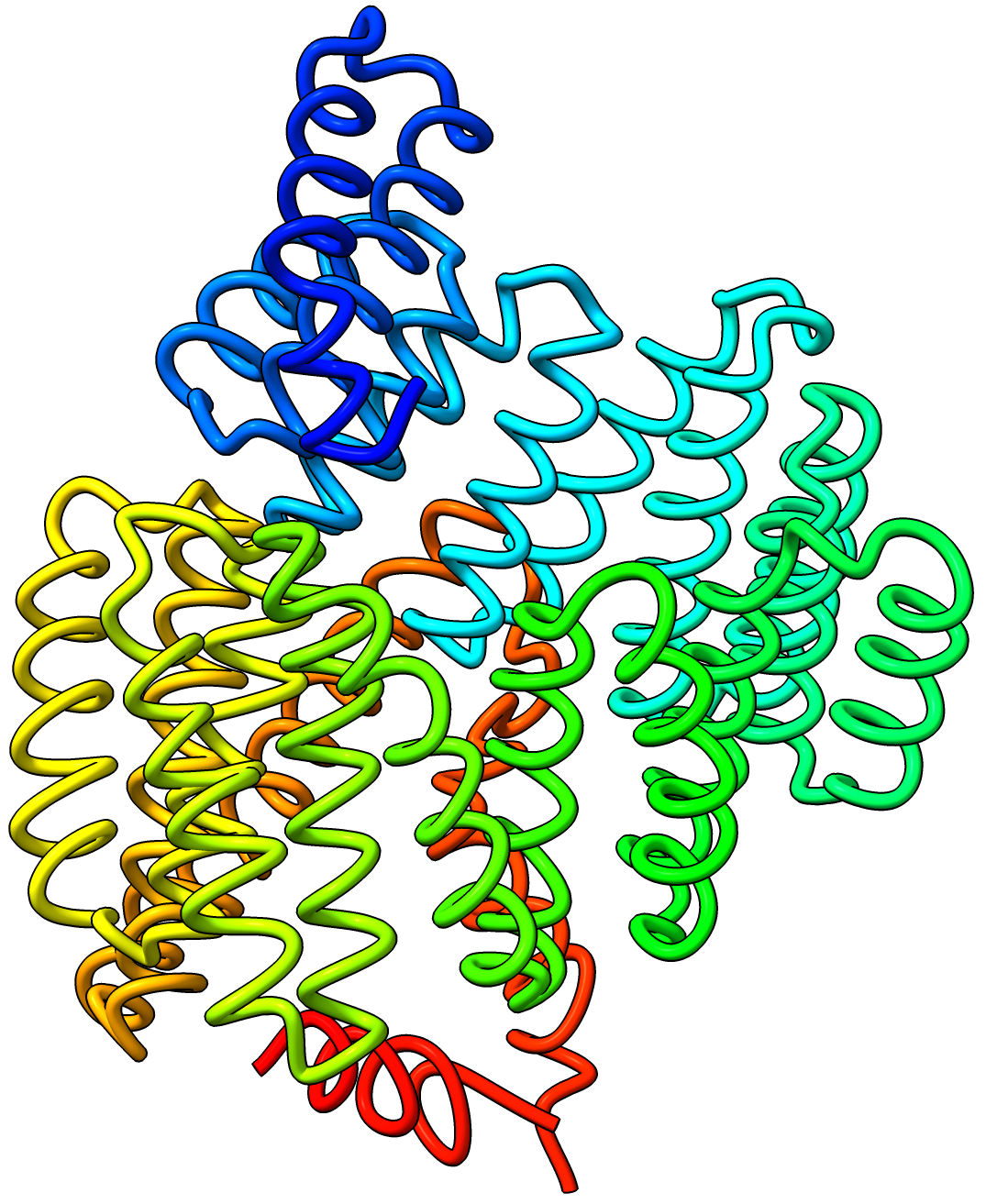}
         \caption{Length: 500.}
         \label{fig:uncond_500}
     \end{subfigure}

     \caption{\textbf{Unconditional co-generation.} (a) pLLDT: 78.4; scTM: 0.64. (b) pLDDT: 67.8; scTM: 0.6. (c) pLDDT: 61.0; scTM: 0.62. (d) pLDDT: 60.6; scTM: 0.60. (e) pLDDT: 77.7; scTM: 0.64.}
     \label{fig:unconditional_proteins}
\end{figure}

We evaluate unconditional co-generation across five length bins $L$ $\in$ \{100, 200, 300, 400, 500\} ($N$=100 proteins each) using standard designability metrics such as scTM, scRMSD, pLDDT, novelty against CATH dataset (as described in Appendix \ref{sec:test_dataset_location}, and sequence recovery (metrics described in Appendix \ref{sec:eval_metrics}). We compare our multi-modal model with ESM3 \cite{esm3} and DPLM2 \cite{dplm2}  and preset our results in \textbf{Table~\ref{tab:uncond_combined_result}} and generated structures in \textbf{Figure~\ref{fig:unconditional_proteins}}. We present the details of evaluation metrics on section \ref{sec:cogeneration_metric}. On average, our model achieves scTM of 0.70, pLDDT of 76.12 and novelty of 0.54, confirming that generated proteins are both structurally plausible and distinct from known CATH dataset.
Our model overpredicts $\alpha$-helices, with an average secondary structure composition ($\alpha/\beta/c$) of 54.6/1.6/43.8\%. This bias is commonly observed when training on computationally predicted datasets such as AFDB and ESMAtlas, and similar trends have been reported in \cite{proteina, kanzi}.
Our pLLDT score of $\geq$ 70 is considered as a physically plausible structure \cite{esm3}, and the agreement between scTM of 0.70 and scRMSD of 9.61 \AA, indicates that generated structures from our model are geometrically consistent with their co-generated sequences. To further validate this consistency, we additionally ran ProteinMPNN~\cite{mpnn} on generated structures and recovered 30.6\% of the original generated sequences, confirming that our structures are designable and that the model has jointly learned the sequence-structure relationship. We find that DPLM-2 achieves strong designability with scTM of 0.92, but it relies on a large pretrained sequence model~\cite{dplm} for initialization and over an order of magnitude more training data. ESM3-Open model which runs the generation process in a cascaded fashion lags behind the other models. These results suggest that \textit{Yeti}'s structure tokens encode information which supports coherent \textit{unconditional co-generation} of protein structure and sequence and also validate \textit{Yeti} as an expressive and efficient structure tokenizer backbone for a multi-modal protein model. We show that discrete, high-utilization token representations enable stable joint generation of protein sequence and structure without large-scale pretraining. 

\begin{table}[t]
    \centering
    \caption{\textbf{Unconditional generation performance.} 
    Reported scores are mean $\pm$ std. Seq Recov means sequence recovery.
    \textbf{Top:} Results across lengths ($N{=}100$ proteins per bin).
    \textbf{Bottom:} Comparison with existing works. DPLM-2 \cite{dplm2} utilizes a pretrained sequence model \cite{dplm} for initialization, and ESM3-Open \cite{esm3} follows a conditional sequence$\to$structure paradigm. Novelty is the average high TMScore in CATH Dataset.} \label{tab:uncond_combined_result}

    \vspace{2mm}
    \small
    \setlength{\tabcolsep}{6pt}

    \vspace{4mm}

    \begin{tabular}{lcccccc}
        \toprule
        Method & Param & scTM ($\uparrow$) &  scRMSD (\AA$,\downarrow$) & pLDDT ($\uparrow$)  & Novelty ($\downarrow$) & Seq Recov ($\%,\uparrow)$\\
        \midrule
        ESM3-Open & 1.4B & $0.46 \pm 0.18$ & $30.63 \pm 19.61 $ & $64.10 \pm 13.70 $ & 0.92 & 28.40 \\
        DPLM-2 & 650M &  $ 0.87 \pm 0.10$ & $3.98\pm 3.52$ & $82.88 \pm 6.57$ & 0.95  & 37.00 \\
        \midrule
        \textbf{Ours} & 224M & $0.70 \pm 0.05 $ & $9.61 \pm 4.28 $  & $76.12 \pm 7.67 $& 0.54 & 30.60 \\
        \bottomrule
    \end{tabular}
\end{table} 

Performance in deep learning scales predictably with model size, data, and compute \cite{scaling}. We therefore expect that scaling our multi-modal model in parameters, dataset size, and training budget will yield substantial gains as seen in Proteína \cite{proteina} and ESM3 \cite{esm3}. The quality of a discrete token-based protein multi-modal model is upper-bounded by the fidelity and expressiveness of the structure tokenization, and the design choices for co-generation remain mostly unexplored \cite{co-gen-review}. A thorough investigation of scaling behavior and learning capabilities in this setting is an important direction and deserves a separate study in itself which we leave to future work; the present results are intended as a proof-of-concept establishing the viability of \textit{Yeti}'s tokenization as a foundation for training such models.

\vspace{-5pt}
\subsubsection{Flow Trajectory Dynamics}
\vspace{-10pt}
\textbf{Q4}. \textit{What are the dynamics of protein folding during reconstruction based on Yeti tokens?}\\
Biologically, protein folding requires a newly synthesized polypeptide to navigate a very large conformational space to settle into its final state. Yet proteins can fold in seconds or less. While the Levinthal Paradox and Kinetic Partitioning suggest proteins sample fast, directed folding pathways \cite{levinthal, levinthal1968there}, the Energy Landscape Theory proposes they descend a funnel-shaped landscape toward a thermodynamically stable final conformation \cite{onuchic1996protein, onuchic1997theory}. However, neither theories clearly define the folding mechanism. Several models have been proposed which offer perspective on the folding mechanism: the Framework Model emphasizes early formation of local secondary structures ($\alpha$-helices, $\beta$-strands) followed by their assembly into tertiary structure \cite{karplus1976protein, anfinsen1973principles}, the Hydrophobic Collapse Model proposes rapid nonspecific compaction into a disordered core preceding final rearrangement \cite{dill1985theory, gutin1995burst}, the Nucleation-Condensation Model posits that a folding nucleus of weak secondary and tertiary interactions forms first, around which the remaining structure condenses \cite{itzhaki1995structure}, and the Nucleation-Propagation Model suggests a small structural nucleus propagates sequentially until the full native fold is reached \cite{wetlaufer1973nucleation}.

While deep learning has transformed protein structure prediction \cite{jumper2021highly}, how structure emerges during folding remains a fundamental open question with implications for molecular dynamics, folding mechanisms, protein function, and related phenomena \cite{chen2023protein}. To better understand the emergence of structure during reconstruction, we examined the folding dynamics throughout the reconstruction process. We note that the flow trajectory represents a mathematical interpolation rather than a physical folding pathway. In this analysis, we focus on the order in which structural features emerge, providing insight into how the \textit{Yeti} decoder operates on the tokenized latent space.

\begin{figure}[t]
    \centering
    \begin{subfigure}[b]{0.495\textwidth}
        \centering
        \includegraphics[width=\linewidth,height=5cm,keepaspectratio]{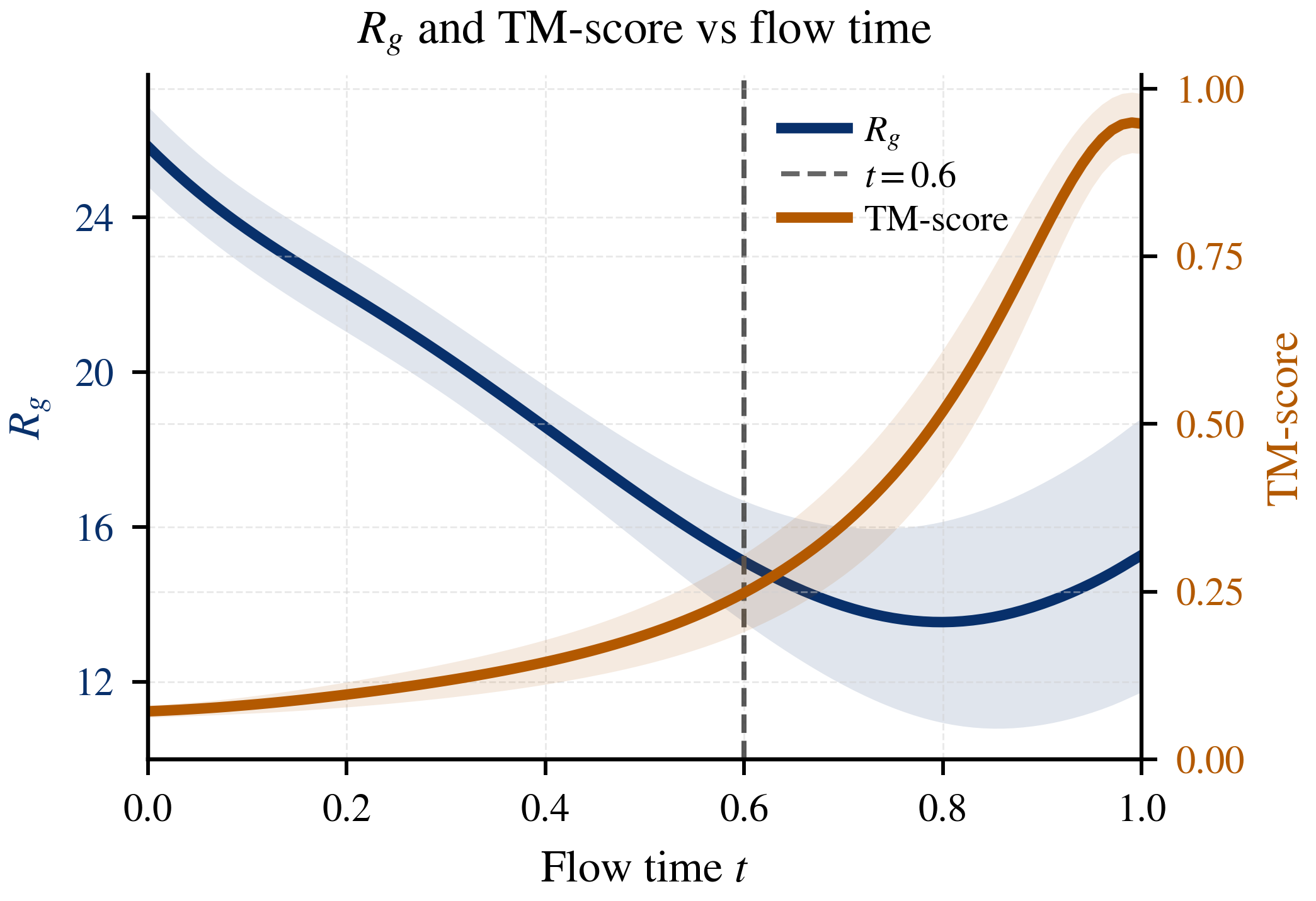}
        \caption{}
        \label{fig:rg_plot}
    \end{subfigure}
    \hfill
    \begin{subfigure}[b]{0.495\textwidth}
        \centering
        \includegraphics[width=\linewidth,height=5cm,keepaspectratio]{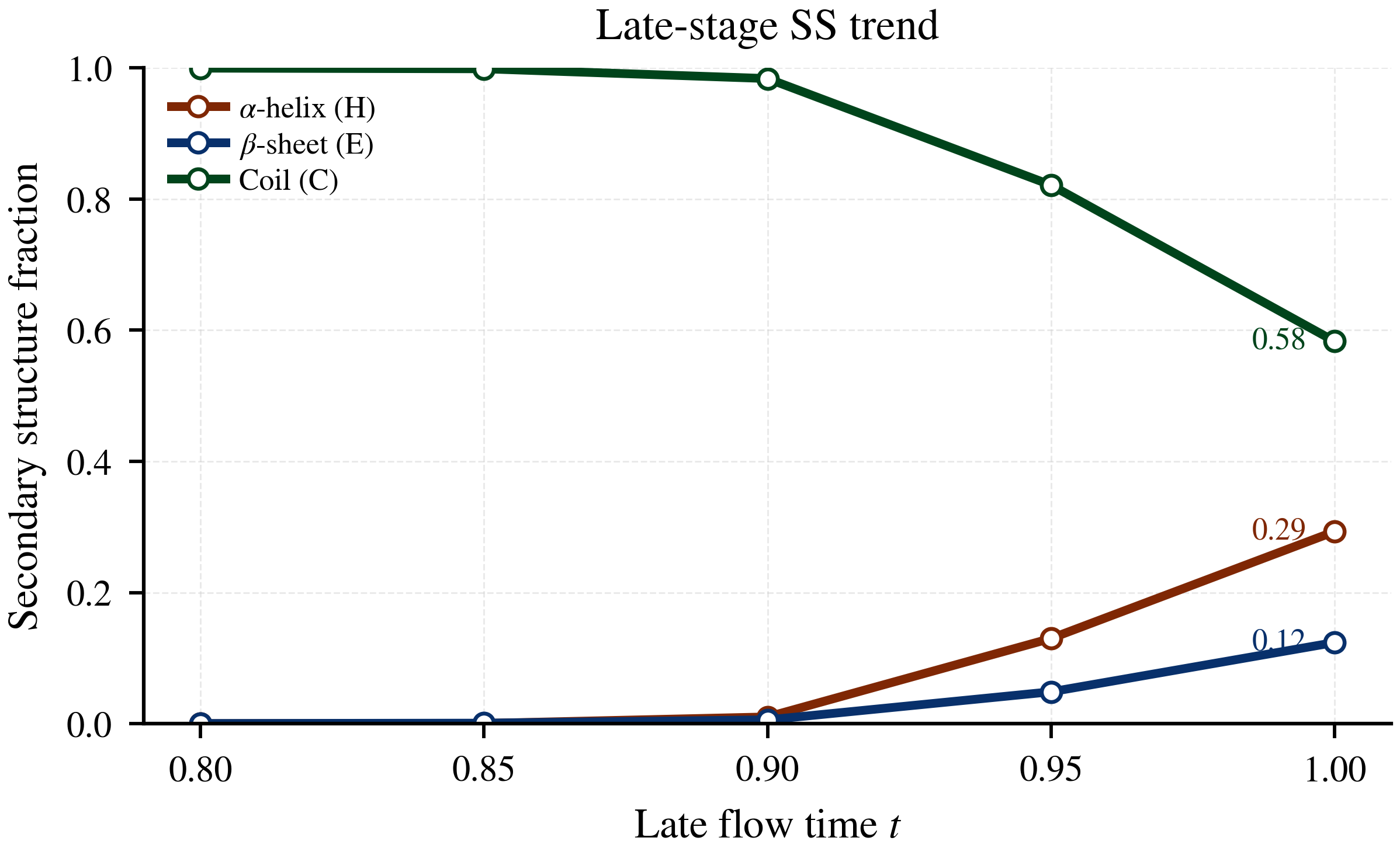}
        \caption{}
        \label{fig:tm_plot}
    \end{subfigure}

    \caption{\textbf{Protein Folding Dynamics during Reconstruction.} (a) The Radius of Gyration ($R_g$) trajectory shows that global compaction begins early ($t < 0.6$), while the TM-score across the flow-matching trajectory continues to increase toward the correct protein topology. The shaded region represents $\pm 1$ standard deviation. (b) Late-stage Secondary Structure (SS) emergence shows the fractions of secondary structure in the final $20\%$ of the flow trajectory.}
    \label{fig:radius_tm}
\end{figure}

Towards this, we visualize the trajectory of the Flow Matching decoder during the reconstruction for diverse 5,000 CATH test dataset. At each step $t \in [0, 1]$, we compute RMSD and TM-score \cite{tmscore} against the known ground truth structure, assign per-residue secondary structure labels using the P-SEA algorithm \cite{sse_annotation_biotite}, and track the radius of gyration, defined as $R_g = \sqrt{\frac{1}{N} \sum_{i=1}^{N} \|\mathbf{r}_i - \bar{\mathbf{r}}\|^2}$,
where $\mathbf{r}_i$ is the $C\alpha$ coordinate of residue $i$ and $N$ is the sequence length. $R_g$ measures how compactly a protein's atoms are distributed around their center of mass where lower values indicate a more tightly packed, well-folded conformations. The trajectory analysis as shown in \textbf{Figure \ref{fig:radius_tm}(a)} demonstrates a clear non-monotonic emergence of structural features during the decoding process. The radius of gyration ($R_g$) indicates that the structures exhibit an early phase where the protein gradually become more compact up to a minimum, and then increase slightly, where as the TM-score increased gradually in the early stage and then accelerated in the late phase. Furthermore, TM-Score exhibited a transition at $t > 0.90$, where secondary structure elements ($\alpha$-helices and $\beta$-sheets) rapidly coalesce \textbf{Figure \ref{fig:radius_tm}(b)}. This suggests the tokenizer effectively decouples coarse-grained density from high-frequency structural details. This preliminary analysis offers a perspective on what generative models may implicitly learn about this process and how the structure emerges during the process.

\section{Conclusions, Limitations \& Future Work} \label{sec:conclusion_limitations_futureWork}
\textit{Conclusions.} The use of Transformer architectures for protein models is a powerful paradigm showing much progress towards diverse down-stream tasks (e.g., reconstruction, generation, etc.,). To fully realize this potential, such models need to ingest diverse data modalities (e.g., sequence, structure, function text) to form an integrated representation of the protein. Core to this approach is the ability to encode the 3D structure of proteins as discrete tokens. Here, we introduce \textit{Yeti} as a compact protein structure tokenizer and demonstrate several favorable properties compared to much larger or more limited models on diverse datasets: \textit{Yeti} had superior codebook utilization and token diversity, highly competitive protein structure reconstruction, strong co-generation capabilities. Finally, we found that the process of Flow Matching based protein folding exhibited two phases. Together, these results demonstrate that \textit{Yeti} will be a powerful tool for training multi-modal protein models at scale. The source code of \textit{Yeti} will be released in GitHub repository.

\textit{Limitations.} Our comparative analysis was limited to methods that showed capabilities for both reconstruction and generation tasks. Training external baselines for generative tasks would require significant computational overhead; thus, we focused our resources on rigorously testing the limits of our tokenizer within a multi-modal context. Furthermore, while we provide a proof-of-concept for \textit{unconditional sequence-structure co-generation} by training both sequence-structure pairs from scratch, we did not perform the same level of exhaustive hyperparameter tuning or benchmarking as seen in specialized models such as  \textit{La-Proteina} \cite{la_proteina}. We anticipate that further scaling of the dataset and refinement of the multi-modal model’s parameters will yield significant performance gains. 

\textit{Future Directions.} The multi-modal can be easily extended into protein folding, inverse sequence design, and co-generation of both modalities specific to a protein function, which has high potential for protein design applications. Incorporation of additional data modalities, such as function annotations, should present no technical problems (e.g., ESM3). Finally, as with essentially all current protein models, we use a single conformation of each proteins structure, based on its crystal structure. Future work should focus on methods and models to incorporate diverse structures into training so models can learn the conformational landscape of proteins.

\begin{ack}

This research used resources of the National Energy Research Scientific Computing Center (NERSC), a Department of Energy User Facility using NERSC GenAI DDR award ERCAP0031977.
\end{ack}



\medskip

{
\small

\bibliographystyle{unsrt}
\bibliography{references}

}






\appendix

\clearpage
\section*{Appendix Contents}

\makeatletter
\@starttoc{apx}
\makeatother
\clearpage

\section{Tokenizer Scaling Analysis} \label{sec:model_scaling}
\addcontentsline{apx}{section}{Tokenizer Scaling Analysis}

\begin{figure}[t]
     \centering
     \hfill
     \begin{subfigure}[b]{0.495\textwidth}
         \centering
         \includegraphics[width=\textwidth]{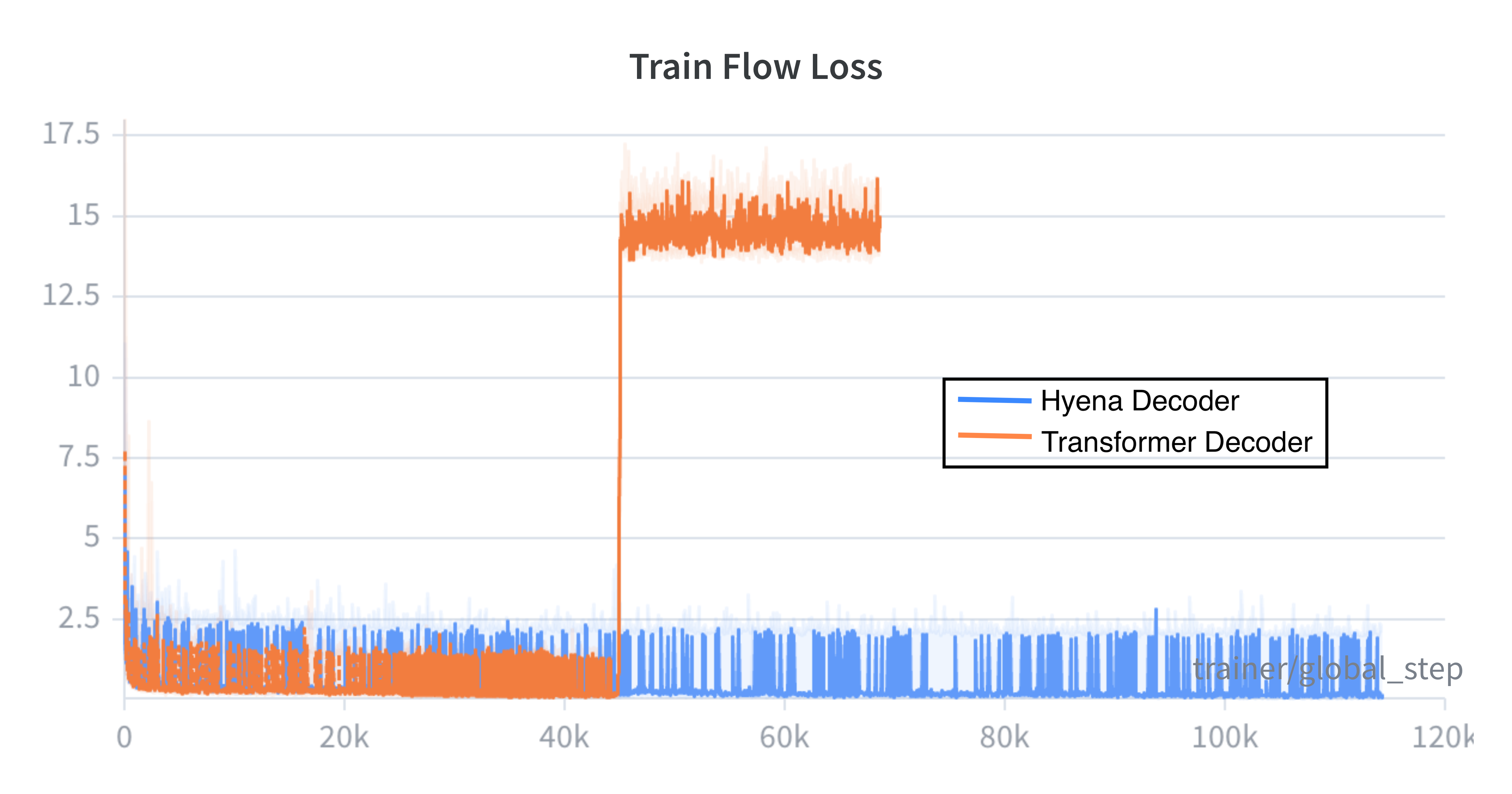}
         \caption{}
         \label{fig:flow_loss_hyena_tranf}
     \end{subfigure}
     \begin{subfigure}[b]{0.495\textwidth}
         \centering
         \includegraphics[width=\textwidth]{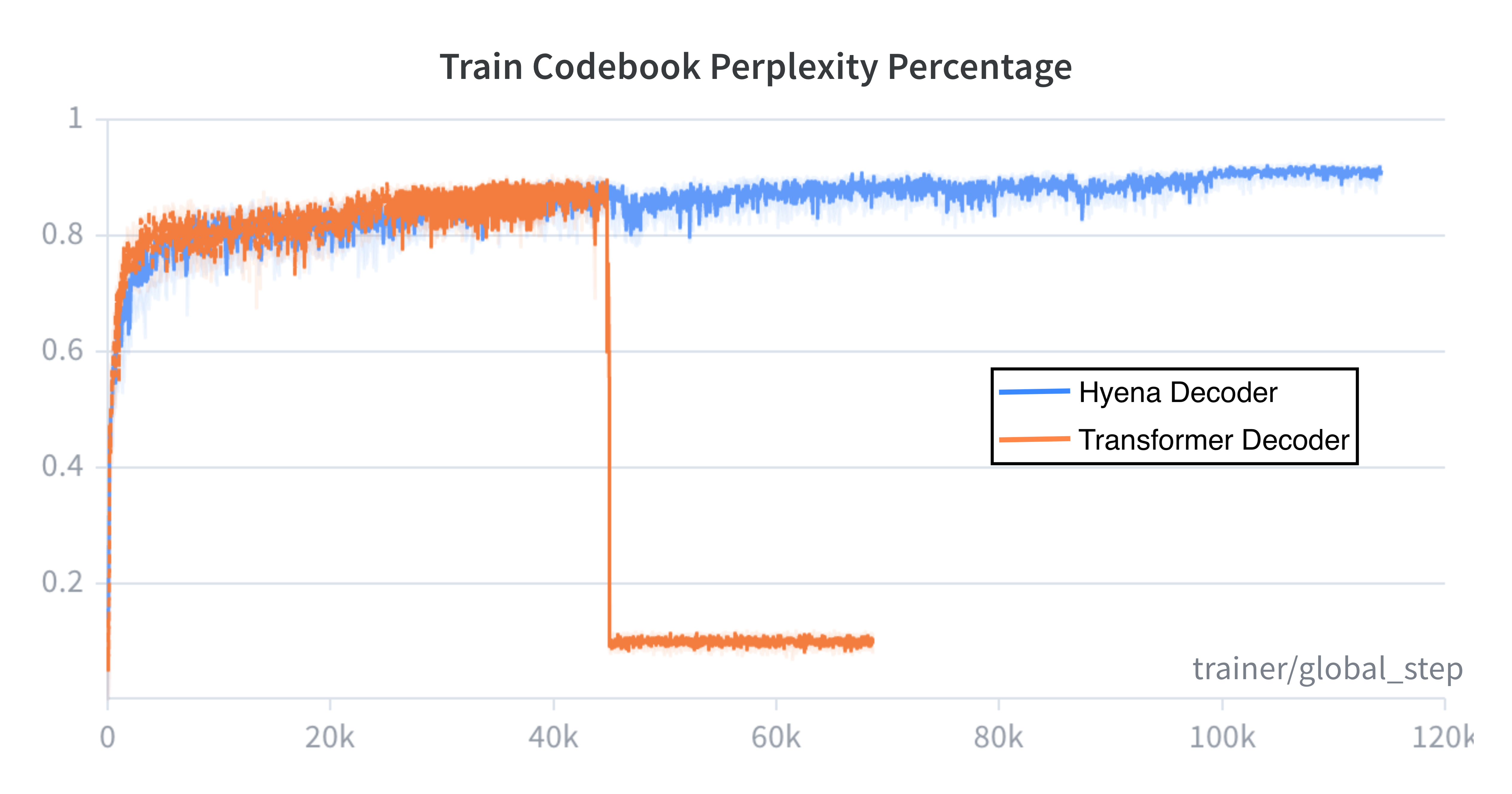}
         \caption{}
         \label{fig:train_codebook_hyena_tranf}
     \end{subfigure}

        \caption{Training curves of StripedHyena-based~\cite{stripehyena} vs.\ Transformer-based~\cite{vaswani2017attention} decoders ($x$-axis: gradient steps). \textbf{(a)} Training flow loss. \textbf{(b)} Codebook perplexity utilization $\in$ (0, 1], where 1 refers to full utilization of all 8,192 codebook entries. The Transformer decoder stagnates after 40K steps, while StripedHyena maintains stable training and higher codebook utilization throughout. Results shown for the \textit{d512-L} configuration (8 encoder / 8 decoder layers, 34.2M encoder parameters). The Transformer-based decoder contains 48.7 M parameters while StripedHyena-based decoder contains 28.3 M parameters.}
        \label{fig:codebook_loss}
\end{figure}

Although scaling Transformer-based generators has been central
to recent advances \cite{gemini, chameleon}, the tokenizer component itself is rarely scaled, leaving open questions about how auto-encoder design choices influence both its objective of reconstruction and downstream generative performance \cite{scaling_visual_tokens_learnings}. To investigate whether protein tokenizer warrant the same scaling efforts as generators, we address two primary bottlenecks : \textit{architecture limitations} and \textit{data scale}. As shown in \textbf{Figure~\ref{fig:codebook_loss}}, early in our development we trained a Transformer-based~\cite{vaswani2017attention} decoder and observed that training stagnated after 40K steps. We later switched to a StripedHyena-based~\cite{stripehyena} decoder which resulted in more stable training. We tested both models using their stable checkpoints on CASP14 test dataset and the StripedHyena-based decoder resulted in better reconstruction quality, as such we selected this as our decoder and  present subsequent ablations based on this decoder.  StripedHyena architecture has been used in Evo \cite{evo, evo2} line of models and a thorough discussion and study is presented in \cite{stripehyena, stripehyena2}.

\subsection{Architecture Analysis} \label{sec:architecture_analysis}
\addcontentsline{apx}{subsection}{Architecture Analysis}

To identify the best \textit{tokenizer' architecture} we trained twelve models with different model configuration on training dataset $\mathcal{D}_{\textit{AFDB}}$. We present the ablation results (reconstruction) in \textbf{Table~\ref{afdb_cfg}} across twelve configurations from three hidden dimensions \textit{d}$\in${128, 256, 512} and four depth levels (2/2, 4/4, 8/8, 16/16 encoder/decoder layers), evaluated on CASP14 proteins with $50 \le L \le 256$, where $L$ is the number of residues present in the  protein.

\begin{table}[h]
  \caption{Ablation on model scaling. \textbf{Bold} and \underline{underline} indicate best and second-best performance, respectively. Per-structure token diversity $\bar{U}$ (more on \ref{sec:rmsd_tmscore}). To isolate the effects of network capacity, the number of attention heads is scaled proportionally to the hidden dimension $d$ to maintain a constant head dimension of $d_{head} = 64$ across all configurations.}
  \vspace{0.2em}
  \label{afdb_cfg}
  \centering
  \small
  \setlength{\tabcolsep}{5.5pt} 
  \begin{tabular}{l l r c c c}
    \toprule
    \textbf{Model} & \textbf{Arch.} (E/D) & \textbf{Params.} (M) & \textbf{RMSD} (\AA, $\downarrow$) & \textbf{TM-Score} ($\uparrow$) & \textbf{$\bar{U}$} ($\%$,$\uparrow$) \\
    \midrule
    \textbf{d128-S}  & 2 / 2   & 1.3   & 2.07  & 0.84  & 90.77  \\
    \textbf{d128-M}  & 4 / 4   & 2.3   & 1.90  & 0.86  & 93.01  \\
    \textbf{d128-L}  & 8 / 8   & 4.2   & 1.87  & 0.87  & 94.92  \\
    \textbf{d128-XL} & 16 / 16 & 8.1   & 1.35  & 0.93  & 96.79  \\
    \addlinespace[0.4em]
    \textbf{d256-S}  & 2 / 2   & 4.7   & 1.85  & 0.88  & 95.32  \\
    \textbf{d256-M}  & 4 / 4   & 8.4   & 1.11  & 0.95  & 97.31  \\
    \textbf{d256-L}  & 8 / 8   & 15.8  & 1.21  & 0.94  & 97.53  \\
    \textbf{d256-XL} & 16 / 16 & 30.7  & 1.18  & 0.94  & 97.49  \\
    \addlinespace[0.4em]
    \textbf{d512-S}  & 2 / 2   & 17.9  & 1.77  & 0.88  & 95.08  \\
    \textbf{d512-M}  & 4 / 4   & 32.8  & 1.16  & 0.94  & \underline{97.78}  \\
    \textbf{d512-L}  & 8 / 8   & 62.5  & \textbf{0.93}  & \textbf{0.96}  & \textbf{98.94}  \\
    \textbf{d512-XL} & 16 / 16 & 121.0 & \underline{1.06}  & 0.95  & 97.45  \\
    \bottomrule
  \end{tabular}
  \vspace{0.2em}
  
  \raggedright
{\footnotesize \textit{Note:} E/D: Encoder/Decoder layers. d$N$: Hidden dimension of size $N$. Attention heads: For d128: $2$ heads ($128 / 2 = 64$). For d256: $4$ heads ($256 / 4 = 64$). For d512: $8$ heads ($512 / 8 = 64$).}
\end{table}

From the \textbf{Table~\ref{afdb_cfg}}, we find that increasing hidden dimension $d$ provides better local reconstruction measured by RMSD, than increasing depth at fixed $d$. We derive this conclusion by analyzing \textit{d256-M} with 8.4M parameters, 4 layers has RMSD of 1.11, which outperforms \textit{d128-XL}  with 8.1M parameters, 16 layers has  RMSD of 1.35, despite the latter having four times as many layers. 
This pattern holds across the Table, where at every depth level, increasing $d$ from 128 to 256 or from 256 to 512 yields consistent RMSD improvements. This is consistent with findings in vision tokenization literature, where wider networks learn rich local feature representations improving reconstruction quality \cite{scaling_visual_tokens_learnings, yao2025reconstruction}, and is also biologically interpretable where wider layers can capture more complex residue-environment relationships such as capturing secondary structure, local geometry, and sequence context simultaneously.

Based on the ablation study presented in \textbf{Table~\ref{afdb_cfg}}, our optimial configuration tokenizer is \textit{d512-L} with 62.5M parameters having 8 layers achieving RMSD of 0.93 \AA, TM-Score of 0.96, and codebook utilization of 98.94\%. We also identify that the larger model \textit{d512-XL} with 121M parameters with 16 layers has RMSD of 1.06 \AA, despite having twice the layers \textit{(8 vs 16)} and nearly twice the parameters. This performance drop likely stems from data limitations rather than architecture deficiencies. The dataset $\mathcal{D}_{\textit{AFDB}}$ with 614,718 samples is sufficient for a $\approx$ 60M parameter model but falls short of the requirements for a $\approx$ 120M parameter model under optimal scaling laws. We anticipate that larger architectures will realize their full potential once paired with larger-scale datasets and longer training.

\subsection{Encoder-Decoder Asymmetry} \label{sec:encoder_decoder_asymmetry}

\addcontentsline{apx}{subsection}{Encoder-Decoder Asymmetry}
To further understand the model configuration, we ran an ablation on Encoder-Decoder layer asymmetry for our optimal configuration model(\textit{d512-L}). \textbf{Table~\ref{tab:layer_asymmetry}} isolates the effect of architecture asymmetry where we fix the total depth at 16 layers but change the symmetry, varying the encoder-decoder split from 14/2 to 2/14.

We find that the symmetric allocation is optimal for our model. The balanced 8/8 configuration achieves the best RMSD of 0.93 \AA \ and TM-score of 0.96, conveying that equal investment in encoding and decoding capacity is the right default at this scale. The Enc-Heavy configuration (12/4) achieves RMSD of 0.99 \AA, which is only slightly worse than the symmetric 8/8, while Dec-Heavy (4/12) increases RMSD to 1.74 \AA. We find this as the encoder must compress rich 3D geometry into a finite discrete codebook vocabulary before the quantization bottleneck which is a task that requires understanding representations of local structure, residue neighborhoods, secondary structure context, and inter-residue geometry. If encoder depth is insufficient, this information is irreversibly discarded at quantization which then leads the discrete tokens which are underspecified, and no amount of downstream decoder capacity can recover what was not captured, as consistent with \cite{scaling_visual_tokens_learnings} where scaling encoders yields gains for either reconstruction or generation. In contrast, the flow matching decoder operates on already-quantized tokens representation and uses its trajectory to recover geometric detail. Because the flow decoder inherently models a distribution over structures consistent with each token, it partially compensates for imprecise encoding which is why we believe the Enc-Heavy (12/4) remains competitive while Dec-Heavy (4/12) does not.

The Enc-Extreme configuration of 14/2 layers has RMSD of 1.24 \AA \ which is not better than Enc-Heavy configuration of 12/4 layers with RMSD of 0.99 \AA, indicating that even the flow matching decoder requires a minimum of four layers for its denoising steps. We summarize these results as a design principle for flow-based protein structure tokenizers where when compute is constrained, investing in encoder depth before decoder depth is a better strategy to achieve better reconstruction accuracy. Since this ablation is mostly geared towards reconstruction, we believe a further ablation on generation capabilities is may provide more insights.

\begin{table}[ht]
  \caption{\textbf{Protein Structure Reconstruction Ablation.} Performance comparison of the optimal model configuration (\textit{d512-L}) for proteins ($50 < L \leq 256$). The notation ($\mathcal{D}_{\textit{A}}$, $\mathcal{D}_{\textit{A+E}}$) denotes training data. $\mathcal{D}_{\textit{A+E}}$ was trained on $\mathcal{D}_{\textit{AFDB}}$ followed by the ESMAtlas dataset $\mathcal{D}_{\textit{ESM}}$. $\mathcal{D}_{\textit{A}}$ was trained only on $\mathcal{D}_{\textit{AFDB}}$.  \textbf{Bold} values indicate the best performance between the two variants. \textbf{CB}: Codebook Size; \textbf{\emph{Util.}}: Per-structure token diversity ($\bar{U}$)}
  \label{tab:reconstruction_supp}
  \centering
  \scriptsize
  \setlength{\tabcolsep}{4pt} 
  \begin{tabular}{l cc | ccc | ccc | ccc | ccc}
    \toprule
    & & & \multicolumn{3}{c}{\textbf{CAMEO}} & \multicolumn{3}{c}{\textbf{CASP14}} & \multicolumn{3}{c}{\textbf{CASP15}} & \multicolumn{3}{c}{\textbf{CATH}} \\
    \cmidrule(lr){4-6} \cmidrule(lr){7-9} \cmidrule(lr){10-12} \cmidrule(lr){13-15}
    \textbf{Model} & \textbf{Params} & \textbf{CB} & RMSD$\downarrow$ & TM$\uparrow$ & Util$\uparrow$ & RMSD$\downarrow$ & TM$\uparrow$ & Util$\uparrow$ & RMSD$\downarrow$ & TM$\uparrow$ & Util$\uparrow$ & RMSD$\downarrow$ & TM$\uparrow$ & Util$\uparrow$ \\
    \textbf{\textit{Yeti} ($\mathcal{D}_{\textit{A}}$)} & 62.5M & 8k & 1.09 & {0.96} & {98.4} & 0.93 & {0.96} & {98.9} & 1.48 & {0.94} & {97.6} & 1.35 & 0.94 & {98.7} \\
    \textbf{\textit{Yeti} ($\mathcal{D}_{\textit{A+E}}$)} & 62.5M & 8k & \textbf{1.05} & \textbf{0.96} & \textbf{99} & \textbf{0.81} & \textbf{0.97} & \textbf{99} & \textbf{1.27} & \textbf{0.95} & \textbf{98.2} & \textbf{1.27} & \textbf{0.95} & \textbf{98.8} \\
    \bottomrule
  \end{tabular}
\end{table}

\subsection{Continued Training with More Dataset} \label{sec:continued_training_dataset}

\addcontentsline{apx}{subsection}{Continued Training with More Dataset}
We also notice that the codebook utilization increases monotonically with model capacity up to \textit{d512-L}, reaching 98.94\%, which is the highest of any configuration. The LFQ objective prevents codebook collapse across all scales, but higher-capacity encoders is able to learn richer per-residue representations that more uniformly distribute across the codebook vocabulary. \textbf{Table~\ref{tab:layer_asymmetry}} conveys this message as the decoder heavy models have slightly lower utilization when compared with encoder heavy models. Additionally, slight utilization drop in \textit{d512-XL} (97.45\%) mirrors the RMSD increase and further supports the interpretation that this model is underutilized relative to its capacity on the available data. Towards, this we continued training our optimal configuration model (\textit{d512-L}) with additional dataset ($\mathcal{D}_{\textit{ESM}}$) for 18,000 steps with batch size of 120 ($\approx$ 1 epoch), resulting in the performance gains detailed in \textbf{Table~\ref{tab:reconstruction_supp}}, addressing the \textit{data scale} bottleneck. Further training for longer steps had diminishing return.

\begin{table}[h]
  \caption{Ablation on Encoder-Decoder layer asymmetry. 
  Total capacity is fixed at 16 layers and hidden dimension $d=512$. 
  \textbf{Bold} indicates best performance. \underline{Underline} indicates 
  second best. \textbf{\emph{Util.}}: Per-structure token diversity ($\bar{U}$) (details on metrics are in \ref{sec:rmsd_tmscore}). CASP14 Test Dataset with $50 < L \le 256$, where $L$ is the number of residues present in the  protein.}
  \label{tab:layer_asymmetry}
  \centering
  \small
  \setlength{\tabcolsep}{4.5pt}

    \begin{tabular}{l l r c c c}
    \toprule
    \textbf{Model} & \textbf{Arch.} (E/D) & \textbf{Params.} (M) & \textbf{RMSD} (\AA, $\downarrow$) & \textbf{TM-Score} ($\uparrow$) & \textbf{Util.} ($\%$,$\uparrow$) \\
    \midrule
    \textbf{Enc-Extreme} & 14 / 2  & 68.4  & 1.24  & 0.94  & 98.23  \\
    \textbf{Enc-Heavy}   & 12 / 4  & 66.4  & \underline{0.99}  & \underline{0.96}  & 98.24  \\
    \textbf{Symmetric}   &  8 / 8  & 62.5  & \textbf{0.93}  & \textbf{0.96}  & \textbf{98.94}  \\
    \textbf{Dec-Heavy}   &  4 / 12 & 58.6  & 1.74  & 0.87  & 97.90  \\
    \textbf{Dec-Extreme} &  2 / 14 & 56.6  & 1.41  & 0.92  & \underline{98.52}  \\
    \bottomrule
  \end{tabular}
  
  \vspace{0.2em}
  \raggedright
  {\footnotesize \textit{Note:} Total depth is held constant at 16 layers to 
  isolate the effect of architecture asymmetry. Symmetric (8/8) corresponds 
  to \textbf{d512-L} in \textbf{Table~\ref{afdb_cfg}.}}
\end{table}

\section{Implementation Details} \label{sec:implement_details}
\addcontentsline{apx}{section}{Implementation Details}

\subsection{Structure Tokenizer} \label{sec:structure_tokenizer}
\addcontentsline{apx}{subsection}{Structure Tokenizer}

We trained sixteen model configurations as detailed in the scaling analysis (Section \ref{sec:model_scaling}). The architecture follows an encoder-bottleneck-decoder framework. The encoder consists of a standard Transformer backbone utilizing Multi-Head Self-Attention (MHSA) with rotary position embeddings (RoPE) \cite{roformer} where the query and keys are transformed by rotation matrices which depend on the relative positions. The bottleneck utilizes Lookup-Free Quantization (LFQ) \cite{lfq} with a codebook size of 8,192. We bridge the transformer latent space and the quantization layer using Multi-Layer Perceptrons (MLPs) for both pre- and post- quantization projections. For the decoder, we utilize a StripedHyena \cite{stripehyena}, optimized via a Flow Matching objective to predict 3D protein structures. We used the default configuration from StripedHyena codebase. 

During training, we consider protein sequences with lengths $L \in [50, 512]$, the ground-truth protein coordinates are mean-centered and scaled by a factor of $0.0667$. Models were trained using the AdamW optimizer with a learning rate of $3 \times 10^{-4}$ and a batch size of 26. We utilized a dropout rate of $0.1$ across all transformer layers. To enable Classifier-Free Guidance (CFG) \cite{cfg} during inference, we randomly dropped the conditioning information with a probability of $0.15$ during training. The LFQ module was configured following the hyperparameters presented in MAGVIT-v2 \cite{lfq} with a commitment loss weight of $0.25$ and an entropy loss weight of $0.1$, utilizing a single codebook.
All experiments were ran on a single NVIDIA A100 GPU. The best configuration model was trained for approximately $120\text{K}$ steps, on 96 hours. We monitored the validation loss throughout training and saved the three best-performing checkpoints; the results reported utilize the checkpoint from step $109.5\text{K}$, which had the minimum validation loss.

\subsection{Masked Diffusion Model Sampling}
We define our \textit{sequence-structure co-generation} as an iterative denoising process over a joint discrete space. Let $\mathcal{S}$ and $\mathcal{X}$ denote the amino acid sequence and structure token vocabularies, respectively. A protein of length $L$ is represented by sequence $\mathbf{s} \in \mathcal{S}^L$ and structure $\mathbf{x} \in \mathcal{X}^L$. Starting from a fully masked state $(\mathbf{s}_T, \mathbf{x}_T) = ([\texttt{M}], [\texttt{M}])^L$, the sampler iteratively generates the sequence and structure tokens over $T$ steps, set to 500. At each step $t \in \{T, \dots, 1\}$, the model predicts the distribution over the masked positions conditioned on the currently unmasked context. To prevent early convergence and allow for error correction, we used a stochastic remasking schedule for the sequence modality where a position $i$ that was generated at step $t+1$ may be independently reverted to the mask token $[\texttt{M}]$ at step $t$ with probability $P_{\text{remask}}^{(t)} = \eta_t \cdot \rho$, where $\rho$ is the base remasking probability and $\eta_t$ is a progress-dependent factor that anneals to zero. Simultaneously, the structure modality follows a confidence based approach . The sampler predicts the structure tokens for all masked positions, but only \textit{keeps} to the highest-confidence predictions while the remaining positions are returned to the masked state according to a \textit{cosine schedule}, which was used to train the model. To manage the high codebook size (8,192) of the structure vocabulary, we apply a nested $\text{Top-}p \ \& \ \text{Top-}k$ filter to the logits, allowing the model to focus on the confident tokens. $\text{Top-}p$ provides a hard threshold for probabilities and $\text{Top-}k$ selects $N$ number of high confidence tokens, eliminating the chance of selecting low probability tokens (long-tail distribution).  

We evaluated three configurations, \textit{coherent}, \textit{balanced}, and \textit{diverse} to understand the trade-off between structure accuracy and diversity. The \textit{coherent} configuration with temperature 0.4, Top-$p$ \& Top-$k$ 0.92 and 96, respectively, resulted in the highest pLDDT and TM-scores, and we report this in \textbf{Table \ref{tab:uncond_combined_result}}. On the other hand, the \textit{diverse} configuration with temperature 0.7, Top-$p$ \& Top-$k$ 0.97 and 192, respectively, generated more diverse structure but the average pLDDT was only 53.8 and a self-consistency TMScore (scTM) of 0.50.

\section{Dataset}\label{sec:dataset_stats}
\addcontentsline{apx}{section}{Dataset}

\subsection{Dataset Locations}\label{sec:test_dataset_location}
\addcontentsline{apx}{subsection}{Dataset Location}

\subsubsection{Test Dataset}
\paragraph{CATH.} We downloaded CATH dataset from \url{https://www.cathdb.info/wiki?id=data:index}. We used  \texttt{S20: Domains are clustered to ensure no two sequences share more than 20\% identity.}

\paragraph{CAMEO.} We downloaded the CAMEO dataset from \url{https://huggingface.co/datasets/genbio-ai/casp14-casp15-cameo-test-proteins}.

\paragraph{CASP14.} We downloaded the CASP14 dataset from \url{https://huggingface.co/datasets/genbio-ai/casp14-casp15-cameo-test-proteins}.

\paragraph{CASP15.} We downloaded the CASP15 dataset from \url{https://huggingface.co/datasets/genbio-ai/casp14-casp15-cameo-test-proteins}.

\subsubsection{Training Dataset}
\paragraph{AlphaFold Database.} We downloaded the AlphaFold Dataset from \url{https://github.com/google-deepmind/alphafold/blob/main/afdb/README.md}.

\paragraph{ESMAtlas Database.} We downloaded the ESMAtlas database from \url{https://github.com/facebookresearch/esm/blob/main/scripts/atlas/README.md}. We downloaded only the high confidence structures from : \url{https://dl.fbaipublicfiles.com/esmatlas/v0/full/tarballs/tm_.60_.70_plddt_.80_.90_00.tar.gz} and \url{https://dl.fbaipublicfiles.com/esmatlas/v0/full/tarballs/tm_.60_.70_plddt_.80_.90_01.tar.gz}. The dataset naming convention for example is; ptm from 0.60 to 0.70 and plddt from 0.80 to 0.90 is named \texttt{tm\_.60\_.70\_plddt\_.80\_.90}.

\subsection{Training Dataset Details}\label{sec:train_dataset_stats}
\addcontentsline{apx}{subsection}{Training Dataset Details}
For both $\mathcal{D}_{\textit{AFDB}}$ and $\mathcal{D}_{\textit{ESM}}$, we filtered the structures to retain only high-confidence predictions with mean pLDDT > 70, and at least 80\% of residues with per-residue pLDDT > 70, coils below 70\%, and sequence length $50 \le L \le 512$. By utilizing these thresholds we ensure that training data reflects well-folded, structurally informative proteins rather than disordered or low-confidence predictions. 

\begin{figure}[t]
     \centering
     \hfill
     \begin{subfigure}[b]{0.49\textwidth}
         \centering
         \includegraphics[width=\textwidth]{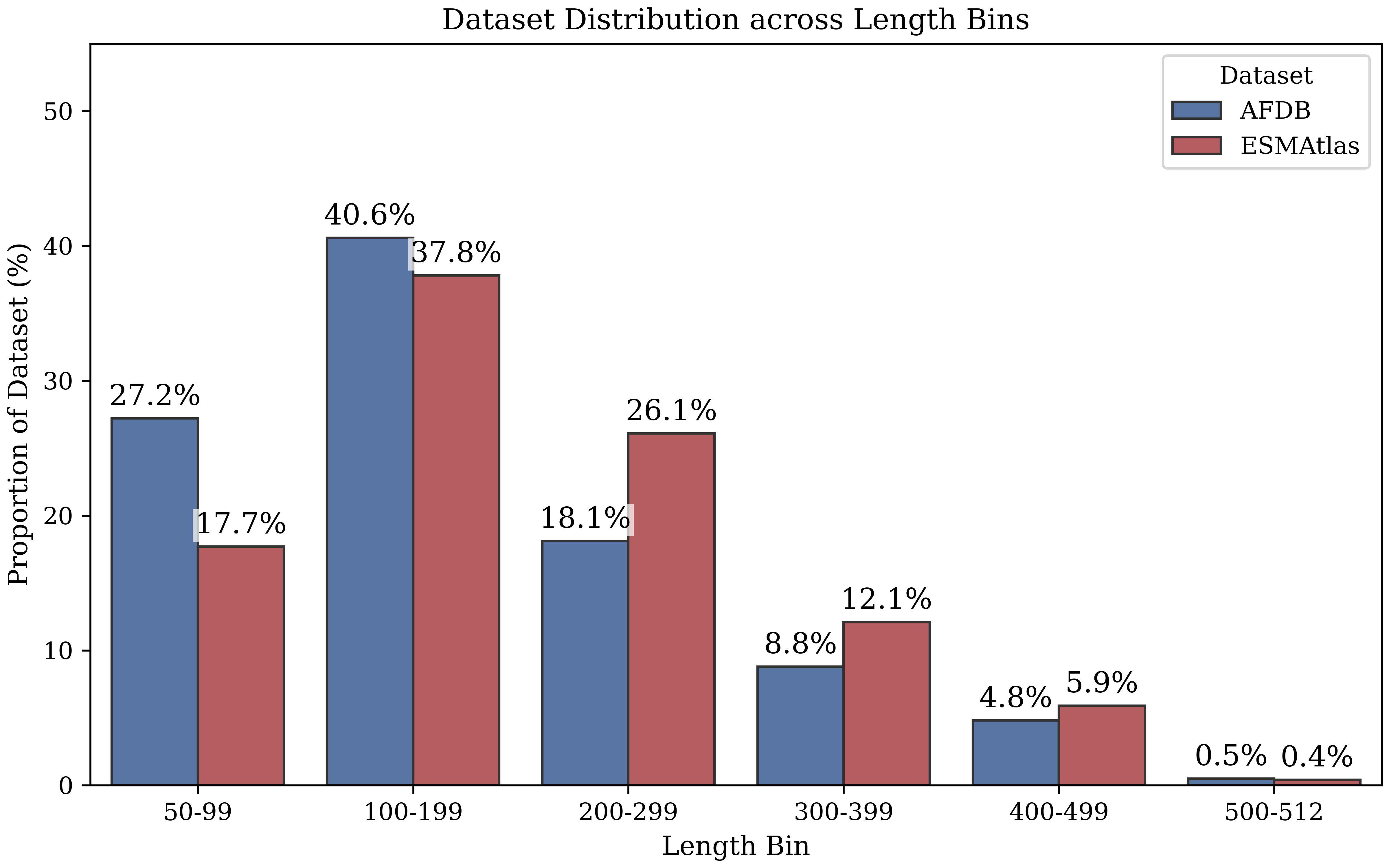}
         \caption{}
         \label{fig:train_dataset_length_dist}
     \end{subfigure}
     \begin{subfigure}[b]{0.49\textwidth}
         \centering
         \includegraphics[width=\textwidth]{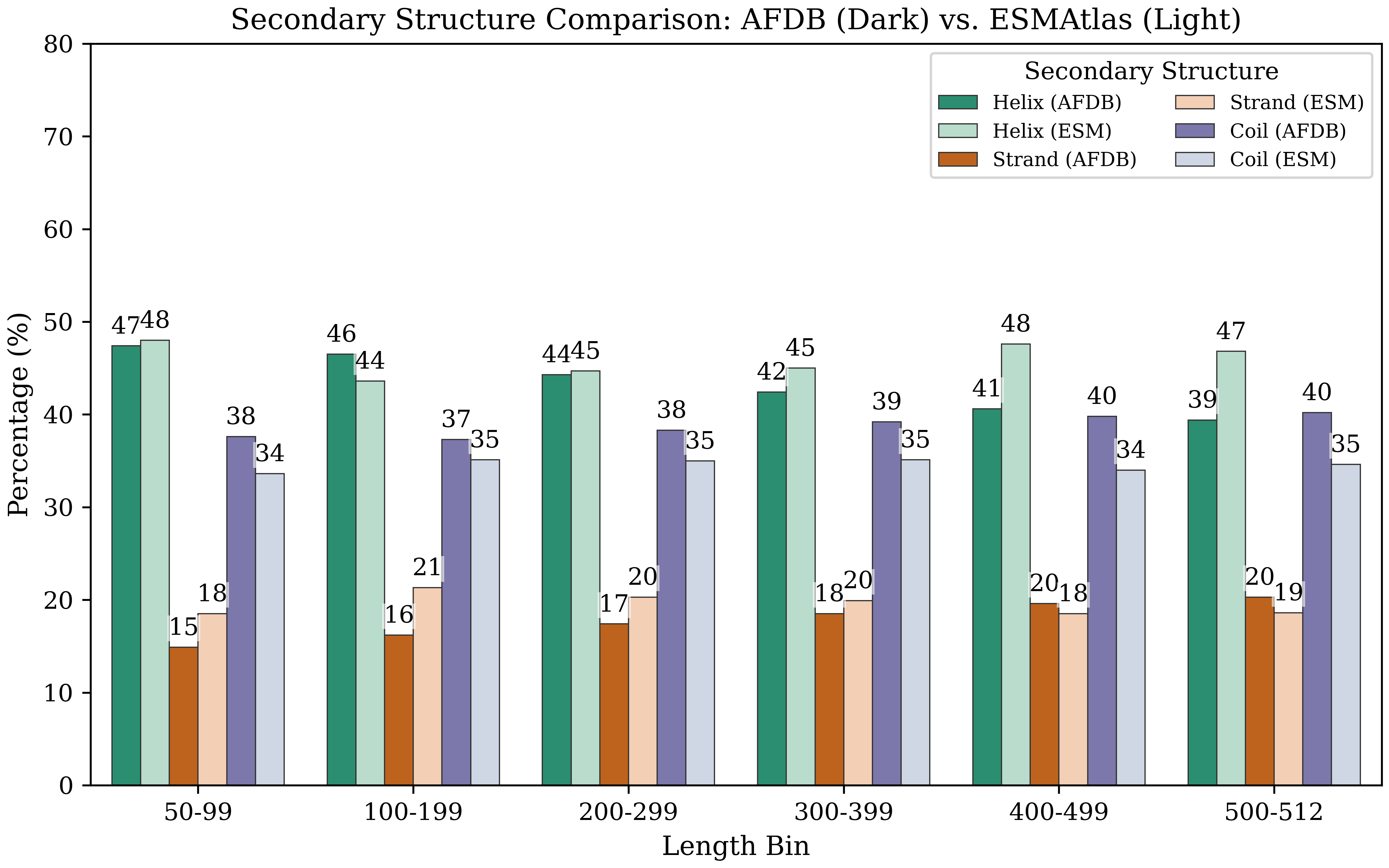}
         \caption{}
         \label{fig:train_dataset_ss_bars}
     \end{subfigure}

        \caption{(a) Proportional distribution of number of protein residues across length bins. (b) Breakdown of $\alpha/\beta$/coils percentages across length bins. Labels show the mean percentage per bin, rounded to the nearest whole number. (c) Mean pLDDT values for AFDB and ESMAtlas across length bins.}
        \label{fig:train_data_composition}
\end{figure}

We present the data statistics in the \textbf{Table \ref{tab:combined_training_data}} and visualize in \textbf{Figure \ref{fig:train_data_composition}}, where we see that the dataset is skewed toward shorter proteins: 67.79\% and 55.42\% of entries fall below 200 residues, for $\mathcal{D}_{\textit{AFDB}}$ and $\mathcal{D}_{\textit{ESM}}$, respectively. Despite this length imbalance, secondary structure composition is stable across bins, helix content decreases gradually from 47.44\% to 39.43\% as length increases, while strand content increases from 14.93\% to 20.34\%, and coil remains near 37-40\% throughout for $\mathcal{D}_{\textit{AFDB}}$. This consistency means that the length distribution does not introduce systematic structural bias into the training distribution, and that the model is exposed to diverse structural contexts at all scales. The average pLDDT (predicted Local Distance Difference Test) scores for $\mathcal{D}_{\textit{AFDB}}$ and $\mathcal{D}_{\textit{ESM}}$ are 87.96, and 90.9, respectively.

\begin{table}[h]
\centering
\small
\caption{Comparative training dataset statistics for AFDB ($\mathcal{D}_{\textit{AFDB}}$) and ESMAtlas ($\mathcal{D}_{\textit{ESM}}$). Values are presented as \textbf{AFDB / ESMAtlas}. Secondary structure (helix, strand, coil) and mean pLDDT are averaged within each sequence length bin.}
\label{tab:combined_training_data}
\setlength{\tabcolsep}{6pt} 
\begin{tabular}{lcccccc}
\toprule
& & & \multicolumn{4}{c}{Secondary Structure (\%)} \\
\cmidrule(lr){5-7}
Length Bin & Proteins & (\%) &pLDDT & Helix & Strand & Coil \\
\midrule
50--99   & 167,279  / 368,065 & 27.2 / 17.7 & 89.0 / 87.7 & 47.4 / 48.0 & 14.9 / 18.5 & 37.6 / 33.6 \\
100--199 & 249,460 / 787,719 & 40.6 / 37.8  & 88.6 / 89.8 & 46.5 / 43.6 & 16.2 / 21.3 & 37.3 / 35.1 \\
200--299 & 111,416 / 544,962 & 18.1 / 26.1 & 88.1 / 92.2 & 44.3 / 44.7 & 17.4 / 20.3 & 38.3 / 35.0 \\
300--399 & 54,152 / 253,227  & 8.8 / 12.1 & 87.7 / 92.2 & 42.4 / 45.0 & 18.5 / 19.9 & 39.2 / 35.1 \\
400--499 & 29,542 / 122,129  & 4.8 / 5.9 & 87.3 / 92.1 & 40.6 / 47.6 & 19.6 / 18.5 & 39.8 / 34.0 \\
500--512 & 2,869 / 9,339 & 0.5 / 0.4 & 87.1 / 91.6 & 39.4 / 46.8 & 20.3 / 18.6 & 40.2 / 34.6 \\
\midrule
\textbf{Total \& Average} & \textbf{614,718 / 2,085,441} & \textbf{100 / 100} & \textbf{88.0 / 90.9} & \textbf{43.4 / 45.9} & \textbf{17.8 / 19.5} & \textbf{38.7 / 34.5} \\
\bottomrule
\end{tabular}
\end{table}

\section{Evaluation Metrics} \label{sec:eval_metrics}
\addcontentsline{apx}{section}{Evaluation Metrics}
To quantify the structure quality, we report the results from following metrics. Together, these metrics helps us to evaluate reconstruction at the local (RMSD), global (TMScore), and codebook utilization, capturing both global usage and per-sequence diversity, providing a overall understanding of tokenizer performance.

\subsection{Evaluation Metrics} \label{sec:rmsd_tmscore}
\addcontentsline{apx}{subsection}{Evaluation Metrics}

\paragraph{Codebook Utilization.} 
To evaluate if the discrete latent space avoids collapse, we assess both global and local diversity. $K$ is codebook size. We compute the global entropy $H = -\sum_{k=1}^K p(k) \log p(k)$ and perplexity = $  \exp(H)$. To capture local variability within individual proteins, we further define the average per-structure diversity $\bar{U} = \frac{1}{N} \sum_{i=1}^{N} \frac{1}{L_i} | \{ z_{i,j} \}_{j=1}^{L_i} |$, which measures the fraction of unique tokens within each structure $i$ of length $L_i$ averaged over $N$ proteins.

\paragraph{TMScore.}
The Template Modeling score  (TMScore) \cite{tmscore} measures the global topological similarity between the reference and ground truth structures. The TMScore is intended as a more accurate measure of the global similarity of full-length protein structures than the RMSD measure. The score $\in [0, 1]$, where 1 indicates a perfect match. A TMScore above 0.5 generally indicates the same fold class; above 0.9 indicates near-identical topology. We installed the TMScore \cite{tmscore} tool into our machine to compute the scores. TMScore tool provides both TMScore and RMSD value. The generation section reports this TMScore and RMSD.

\paragraph{RMSD.}
The Root Mean Square Deviation (RMSD) measures the average Euclidean distance between aligned backbone C$\alpha$ atoms after superposition using the Kabsch algorithm. It is sensitive to local atomic deviations and penalizes small positional errors uniformly, making it the primary measure of fine-grained geometric reconstruction. Lower values indicate more accurate local reconstruction. The reconstruction section reports this RMSD and TMScore computed from TMScore \cite{tmscore} tool.

\subsection{Co-generation Metrics} \label{sec:cogeneration_metric}
\addcontentsline{apx}{subsection}{Co-generation Metrics}

Following the literature \cite{la_proteina, multiflow, dplm2}, we quantify generation quality using four complementary criteria: \textit{foldability}, \textit{designability}, \textit{novelty}, and \textit{sequence recovery}. 

\paragraph{Foldability.} 
Foldability measures whether a generated sequence folds into a physically plausible structure. For each generated structure ($n$=100 per length bin $\in$ [100, 200, 300, 400, 500]), we fold  the generated sequence using ESMFold \cite{esmfold} and report the predicted Local  Distance Difference Test (pLDDT), a per-residue confidence score $\in [0, 100]$ where higher values indicate a well-folded structure.

\paragraph{Designability.} 
Designability measures whether the generated structure and sequence are mutually consistent, that is, whether the generated sequence would naturally fold into the generated structure. Since our model generates both modalities simultaneously, this is evaluated as a self-consistency check where we fold the generated sequence with ESMFold \cite{esmfold} to obtain a predicted structure, then compare it to the structure generated by our model. We report self-consistency TM-score (scTM) between the two structures. Following strict statistics of structures in the PDB \cite{pdb}, scores below 0.17 correspond to randomly chosen unrelated proteins whereas structures with a score $\geq$ 0.5 assume generally the same fold in SCOP/CATH \cite{tmscore}. A high scTM score indicates that the model has learned a joint distribution over sequence and structure that respects their mutual compatibility, rather than generating each modality independently.

\paragraph{Novelty} 
Novelty measures the structure similarity between generated structure and a reference dataset, where lower TM-scores (ranging from 0 to 1) indicate greater novelty. We define the novelty of a generated structure as its maximum TM-score \cite{tmscore} relative to a reference set of 11,752 structures from the CATH dataset (clustered at 20\% sequence identity). For each co-generated structure, we identify its closest structure neighbor in the reference dataset using TMScore and report the average TMScore for that length bin.

\paragraph{Sequence Recovery} 
Sequence recovery is defined as the percentage of amino acid residues in a designed sequence that are identical to the reference sequence at corresponding positions. We utilized ProteinMPNN \cite{mpnn} in default mode to generate eight sequences for each designed structure, and report the maximum recovery score. This metric serves as a self-consistency check to evaluate whether the generated structure can be successfully inverse-designed back to a sequence, ensuring sequence-structure compatibility.

\section{Multimodal Protein Models} \label{sec:multimodal_protein_models}
\addcontentsline{apx}{section}{Multimodal Protein Models}

Recent advancements in \textit{structure-sequence co-generation} have highlighted several architecture trade-offs. ProteinGenerator \cite{ProteinGenerator} performs sequence-space diffusion based on RoseTTAFold \cite{rosettafold} that simultaneously generates protein sequences and structures, enabling guided design of proteins with specific attributes and multifunctional properties. Multiflow \cite{multiflow} utilizes a multimodal flow-matching approach; however, despite its impressive structure generation capabilities, it struggles to generate structurally compatible sequences. Consequently, it resort to instance-level distillation from ProteinMPNN \cite{mpnn} and exhibits subpar performance in folding tasks for given sequences. DPLM-2 \cite{dplm}, a 650 million parameter model \cite{dplm2} leverages a pretrained protein sequence-based language model (pLM)  via warm-up fine-tuning. While this allows the model to inherit strong sequence priors, it remains rooted in a sequence-first paradigm. Similarly, ESM3 \cite{esm3}, a 1.4 to 98 billion parameters model adopts a sequence-first architecture, generating modalities in a cascaded, step-wise manner. The GFP design pipeline of ESM3 alternates between generating structure tokens based on functional motifs and then optimizing the sequence based on the structure, repeated for multiple cycles. ProDiT (Protein Diffusion Transformer) \cite{prodit}, a 321M and 576M parameters variant model with architecture closely based on design of Diffusion Transformer (DiT), models sequence and structure with diffusion processes in their respective state spaces without tokenizing the structures into discrete vocabulary and achieves good performance over co-generation, the code is yet to be released. PLAID (Protein Latent Induced Diffusion), a 2B and a 100M variant DiT \cite{dit} model, performs diffusion directly within the latent space of a pretrained protein folding model ESMFold\cite{esmfold}, which allows it to repurpose the features from pLM for generation task. \textit{La-Proteina} \cite{la_proteina}, a $\approx$ 295M (167 $+$ 128M) parameter model, first trains a conditional VAE, with its encoder mapping input proteins (sequence and structure) to latent variables, and its decoder reconstructing complete proteins from the latent variables and $C\alpha$ coordinates. Leveraging the VAE, it then trains a flow matching model to learn the joint distribution over latent variables and coordinates of the $C\alpha$ atoms and is trained on 46M sequence-structure pairs for atomistic protein design. SaProt \cite{saprot}, a 650 million parameters model introduces the concept of a \textit{structure-aware vocabulary} that integrates residue tokens with structure tokens. The structure tokens are derived by encoding the 3D structure of proteins using Foldseek \cite{foldseek}, Foldseek encodes structures as sequences over the \textit{20-state} 3Di alphabet for fast and sensitive comparison of large structure sets. This has been useful for clustering AlphaFold and ESMAtlas dataset. SaProt utilizes the \textit{SA}-token protein sequence as input, and train a structure-enhanced PLM using the ESM \cite{esm2} backbone on 40 million proteins. More recently, ProCyon \cite{procyon}, a 11 billion parameter multimodal model for protein phenotypes, utilized Llama-3-8B \cite{llama3}, ESM-2 3B \cite{esm2}, and GearNet \cite{gearnet} for its respective modalities. However, this reliance on a heterogeneous assembly of independently pretrained backbones characterizes a late-fusion paradigm. Such an approach precludes the learning of a natively unified representation, as the model lacks a \textit{de novo} joint-training objective for sequence and structure.

In the broader context of multimodal learning, architectures are typically categorized into early, late, or hybrid fusion strategies. Late-fusion models, like ProCyon, integrate high-level features only at the final stages of processing. In contrast, early-fusion mechanisms which initially studied in biological neural networks where sensory inputs are combined at the earliest layers of processing \cite{multisensory, multisensory2}, have been adopted by LLMs such as Gemini \cite{gemini} and Chameleon \cite{chameleon}. Inspired by these developments, we lean on adopting an early-fusion approach, allowing the model to learn the intrinsic, biophysical correlations between sequence and structure from the start of the training process.

\newpage

\end{document}